\documentclass[%
 twocolumn,
nofootinbib,
 amsmath,amssymb,
 aps,
]{revtex4-2}

\usepackage{graphicx}
\usepackage{dcolumn}
\usepackage{bm}
\usepackage{framed}
\usepackage{hyperref}
\usepackage{braket}
\usepackage{dsfont}
\usepackage{color}
\usepackage[dvipsnames]{xcolor}
\begin{document}
\title{Time Series Quantum Reservoir Computing with Weak and Projective Measurements}%
\author{Pere Mujal}
\affiliation{%
IFISC, Institut de F\'{\i}sica Interdisciplin\`{a}ria i Sistemes Complexos (UIB-CSIC),
\\ UIB Campus, E-07122 Palma, Mallorca, Spain
}%
\author{Rodrigo Mart\'inez-Pe\~na}
\affiliation{%
IFISC, Institut de F\'{\i}sica Interdisciplin\`{a}ria i Sistemes Complexos (UIB-CSIC),
\\ UIB Campus, E-07122 Palma, Mallorca, Spain
}%
\author{Gian Luca Giorgi}
\affiliation{%
IFISC, Institut de F\'{\i}sica Interdisciplin\`{a}ria i Sistemes Complexos (UIB-CSIC),
\\ UIB Campus, E-07122 Palma, Mallorca, Spain
}%
\author{Miguel C. Soriano}
\affiliation{%
IFISC, Institut de F\'{\i}sica Interdisciplin\`{a}ria i Sistemes Complexos (UIB-CSIC),
\\ UIB Campus, E-07122 Palma, Mallorca, Spain
}%
\author{Roberta Zambrini}
\affiliation{%
IFISC, Institut de F\'{\i}sica Interdisciplin\`{a}ria i Sistemes Complexos (UIB-CSIC),
\\ UIB Campus, E-07122 Palma, Mallorca, Spain
}%

\date{\today}
\begin{abstract}
Quantum machine learning represents a promising avenue for data processing, also for purposes of sequential temporal data analysis, as recently proposed in quantum reservoir computing (QRC). The possibility to operate on several platforms and noise intermediate-scale quantum devices makes QRC a timely topic. A challenge that has not been addressed yet, however, is how to efficiently include quantum measurement in realistic protocols, while retaining the reservoir memory needed for sequential time series processing and preserving the quantum advantage offered by large Hilbert spaces. In this work, we propose different measurement protocols and assess their efficiency in terms of resources, through theoretical predictions and numerical analysis. We show that it is possible to exploit the quantumness of the reservoir and to obtain ideal performance both for memory and forecasting tasks with two successful measurement protocols. One consists in rewinding part of the dynamics determined by the fading memory of the reservoir and storing the corresponding data of the  input sequence, while the other employs weak measurements operating online at the trade-off where information can be extracted accurately and without hindering the needed memory.
Our work establishes the conditions for efficient protocols, being the fading memory time a key factor, and demonstrates the possibility of performing genuine online time-series processing with quantum systems.
\end{abstract}
%
\maketitle
\section{\label{sec:intro}Introduction}
The availability of big data analyzed and exploited with machine learning methods is one of the main traits of the present information era and the search for enhanced data processing capabilities is spurring research also in quantum approaches~\cite{biamonte2017quantum,cerezo2021variational,RevModPhys_NISQ2022}. Recent quantum machine learning implementations have been reported ranging from NMR platforms \cite{li2015experimental} to noisy quantum computers \cite{havlivcek2019supervised,Peters2021}, and many proposals \cite{cai2015entanglement,hu2019quantum,yao2017quantum,tacchino2019,Liu2021} promise an advantage in the performance with respect to classical approaches. Still, dealing with quantum systems for machine learning poses new challenges not only to preserve fragile quantum coherence from the action of the environment \cite{breuer2002theory} but also to efficiently access the processed information through measurement \cite{wiseman2009quantum}, being these related issues.
In particular, a first distinctive aspect of quantum measurements is that expectation values require an ensemble of copies of the system. Indeed, in most quantum technologies including computation \cite{arute2019quantum}, sensing \cite{Pirandola2018,RevModPhys.89.035002}, or communications \cite{Gisin2007,Munro2012,Chen2021}, different approaches to optimize and reduce the number of measurements have been recently reported \cite{cramer2010efficient,elben2022randomized} also including classical machine learning tools \cite{hentschel2010machine,garcia2021learning,torlai2018neural,palmieri2020experimental}. A second distinctive aspect to deal with is that a measurement performed on a quantum system generally alters its state \cite{skinner2019measurement,li2018quantum,elouard2017role,manzano2021quantum}. This measurement back-action has deep consequences when monitoring a system, beyond a single input-output setup and has been addressed for instance in feedback control \cite{wiseman2009quantum} or in quantum error correction  \cite{Wineland2004,PhysRevX.11.041058}. Our goal here is to introduce a quantum measurement framework for temporal series processing considering both accuracy and back-action.

Many prominent tasks in the context of (classical) machine learning deal with time-series processing, like speech recognition, stock  market forecasting or climate prediction. In contrast to classification tasks \cite{jordan2015machine}, temporal ones require continuous data monitoring and the temporal order within the data sequence to be processed plays a key role \cite{jaeger2004harnessing}. 
Among classical approaches to time-series processing \cite{lukovsevivcius2009reservoir,salinas2020deepar,hewamalage2021recurrent,gauthier2021next}, reservoir computing \cite{Book_Nakajima_Fischer2021} has been successful in recognizing spoken words \cite{larger2017high,Romera2018} and human activity~\cite{Palumbo2016} or in forecasting chaotic dynamical systems \cite{pathak2018model,moon2019temporal}, with appealing features such as easy training and energy efficiency. In this scheme, data are continuously injected into a reservoir system, processed, and extracted, without requiring an external memory avoiding the von Neumann bottleneck \cite{sebastian2020}. 
The reservoir itself can be either an artificial  recurrent neural network \cite{jaeger2001echo} or even a physical substrate \cite{tanaka2019recent}, providing a high dimensional phase space, a non-linear input-output transformation and a fading memory \cite{konkoli2017reservoir}.
\begin{figure*}[ht!]
    \centering
    \includegraphics[width=2\columnwidth]{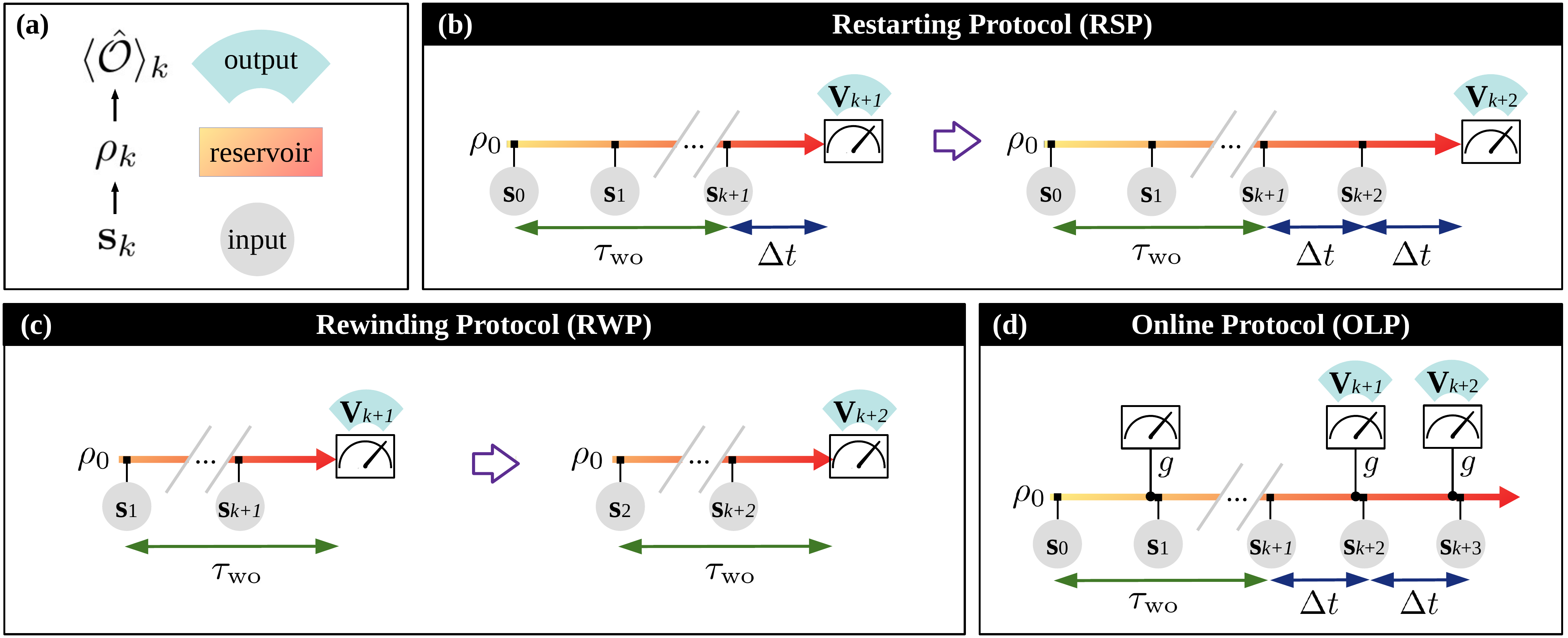}
    \caption{(a) Schematic representation of QRC, in input, reservoir and output layers.  (b)--(d) Measurement restarting, rewinding ond online protocols (RSP, RWP and OLP) for consecutive measurements at time steps $k+1$ and $k+2$ in one ensemble realization.
    For the RSP (b), at each output time step, all the sequence processing is repeated ($N_t$ time steps to gather the output $k+1$,  $N_t+1$ time steps to gather $k+2$, etc.). For the RWP (c),
    $N_{\rm wo}$ steps of the sequence are repeated for each output time step. For the RWP the choice of the reset state at $t_k-\tau_{\rm wo}$ has no effect on the output (here we set it to $\rho_0$). For the OLP with weak measurement (d), the evolution is continuously monitored without repeating any part of the reservoir evolution.
}
    \label{fig:protocols}
\end{figure*}

Recently, reservoir computing has been extended to the quantum regime considering several models ranging from qubits to fermions and bosons, modeling photonic, atomic and solid-state platforms \cite{mujal2021opportunities,memristors_walther}.
Theoretical proposals display successful performances of quantum reservoir computing (QRC) in genuine temporal tasks \cite{Nakajima2017,tran2021learning,chen2020temporal,martinez2021dynamical,nokkala2021gaussian,suzuki2022natural} and in generalization and classification ones \cite{khan2021physical,govia2021quantum,ghosh2019quantum,ghosh2019quantum2,ghosh2021realising,memristors_walther,condmat7010017}, an approach known as extreme learning machine. 
A first experimental implementation of a static  classification task has been realized with NMR of a nuclear spin ensemble in a solid  \cite{negoro2018machine}, while the first exploration of temporal series processing with QRC was recently reported on a quantum digital computer \cite{chen2020temporal}. The potential of exploiting present NISQ devices \cite{Preskill2018quantumcomputingin,RevModPhys_NISQ2022} is indeed one of the motivations for QRC \cite{Nakajima2017}. Still, in order to establish the most promising avenues towards quantum approaches a daunting question is how to \textit{efficiently} monitor and extract \textit{meaningful} information in such experimental platforms, when including the effect of quantum measurement. The scope of this work is to address this point by proposing different measurement protocols and to identify the most promising ones  for implementation, also addressing the crucial issue of the required resources.

Continuous time-series processing needs to be not only efficient in terms of time and energy resources (repetitions, ensemble size) but also reliable in spite of measurement back-action, being these often competing requirements.  The presence of noise due to finite ensembles is common to most quantum technologies and has also been considered in proposals of QRC  \cite{nokkala2021high,govia2021quantum,GhoshReconstructing2021,khan2021physical}.
In time-series processing, the back-action of measurement for continuous monitoring  (i.e. introducing the measurement map for each processing time step) becomes a further crucial factor for realistic implementations, still to be addressed. Here, we will explore protocols based either on repeating (part of) the reservoir dynamics or on weak measurements~\cite{RevModPhys.82.1155,brun2002simple}, known to be noisier and less informative than projective ones, but with the advantage to perturb less the state, therefore preserving time information in the system.
In this work, we analyze the effect of quantum measurement for realistic QRC, assessing the required resources for temporal series processing (considering both finite ensemble and back-actions issues) with analytical and numerical methods (Sec.~\ref{sec:2}). Considering different protocols (Sec.~\ref{sec:prot}) based on
both projective and weak measurements  (Sec.~\ref{numresA}) and taking into account their accuracy (Sec.~\ref{sec:accu}), we will identify successful strategies leveraging performance and resource efficiency. In Sects.~\ref{sec:STM} and ~\ref{sec:santafe}, we illustrate these aspects for time series processing in  tasks requiring respectively memory or forecasting capabilities. Conclusions and future perspective are discussed in Sec.~\ref{sec:conclusions}.
\section{RESULTS}\label{sec:2}
\subsection{Restarting, rewinding and online protocols}  \label{sec:prot}

All QRC architectures \cite{mujal2021opportunities} for time series processing proposed so far share the common feature of information extraction based on quantum measurement at the output layer, Fig.~\ref{fig:protocols}(a). In this section, we present three different measurement protocols of a quantum system driven by a sequence of (classical or quantum) inputs, as illustrated in Fig.~\ref{fig:protocols}(b)--(d), being most considerations generally valid for any monitored quantum system, also  beyond QRC. 
We formally describe the evolution of a  quantum system state in discrete time steps, with $k=t/\Delta t$ labeling the input injection sequence, by means of the recurrence relation
\begin{equation}
\label{eq:rhokunperturbed}
    \rho^{\textnormal{u}}_k=\mathcal{L}_k[\rho^{\textnormal{u}}_{k-1}]
\end{equation}
for the \textit{unperturbed} (u) state dynamics, i.e., in the absence of any measurement process. 
The superoperator $\mathcal{L}_k$ can account both for a unitary evolution as well as for the interaction with an additional external environment.
The input,  $\mathbf{s}_k$ in Fig.~\ref{fig:protocols}, can be introduced either through the sequential update of the state of part of the system state \cite{Nakajima2017} or by means of  a control Hamiltonian or dynamical parameter \cite{govia2022nonlinear,chen2020temporal}.

In an ideal case, at each time step, information is extracted through the expectation value of any observable operator $\hat{\mathcal{O}}$ over an infinite ensemble
\begin{equation}
\label{eq:ideal_means}
    \langle \hat{\mathcal{O}} \rangle^{\infty}_{\rho_k^{\rm u}} \equiv \textnormal{Tr}\left(\hat{\mathcal{O}}\rho^{\textnormal{u}}_k \right).
\end{equation}
In any implementation, however, the expectation value can be inferred only from a finite number,  $N_{\rm meas}$,  of identical copies of the quantum system and the estimated value suffers from a statistical error that represents a first major experimental limitation. 
The second issue of paramount relevance is represented by the disturbance introduced by a quantum measurement, which modifies the ``unperturbed" dynamics of Eq.~(\ref{eq:rhokunperturbed}) at each time step $k$.  
A common way to circumvent this back-action issue, --where the effect of quantum measurement is not accounted for-- is through the reinitialization of the dynamics after each measurement, as  usually implicit in QRC proposals so far \cite{mujal2021opportunities}. 
As a drawback, one would need to repeat the experiment $N_{\rm meas}$ times for each of the time steps $t_k$ of the input time series to be processed.  This scheme, which we will refer to as the \textit{restarting} protocol (RSP), is sketched in Fig.~\ref{fig:protocols}(b). It was used in the quantum digital computer implementation in Ref. \cite{chen2020temporal}.

A significant improvement with respect to the RSP can be attained in reservoir computing considering its dynamical features, and in particular the echo state property, that is its ability to process time series independently on the initial state $\rho_0$ of the reservoir \cite{grigoryeva2018echo}.
We introduce the washout time $\tau_{\rm wo}$, as the time necessary for the dynamical system to lose any information from initial conditions, so that the reservoir computer depends effectively on the input recent history, a property also known as fading memory (Fig.~\ref{fig:protocols}(b)).
This suggests that a more efficient strategy than the RSP is to limit the repetition time to the shorter washout time,  restarting the dynamics from $t_k-\tau_{\rm wo}$ instead of $t_0$ \cite{chen2020temporal}. 
This \textit{rewinding} protocol (RWP) is devised to  optimize the resources needed to overcome the measurement feedback by re-initialization and is characterized by a sliding repetition time interval $\tau_{\rm wo}$ with an associated number of input injections $N_{\rm wo}\equiv \tau_{\rm wo}/\Delta t$ 
(Fig.~\ref{fig:protocols}(c)).  

In both RSP and RWP, the measurement outcomes, $\mathbf{V}_{k\,l}$, at each time step $k$ and for several realizations $l=1,\,...,N_{\rm meas}$, allow one to approximate  the expectation value of the observables as an average over a large number of measured values $\langle \hat{\mathcal{O}} \rangle^{N_{\rm meas}}_{\rho_k^{\rm u}}$ with a precision proportional to $1/\sqrt{N_{\rm meas}}$.
In other words, the statistical uncertainty limits the approximation to the ideal expectation values as  $\langle \hat{\mathcal{O}} \rangle^{\infty}_{\rho_k^{\rm u}}=\langle \hat{\mathcal{O}} \rangle^{N_{\rm meas}}_{\rho_k^{\rm u}}+O\left(N^{-\frac{1}{2}}_{\rm meas}\right)$. 
These protocols are designed so that the measurement back-action does not need to be taken into account but require to repeat,  for each input injection, part or all of the processing sequence, requiring then to store these inputs in an external memory. An important question arises in this context, \textit{is there a way to avoid restarting (or rewinding) the process after each output extraction?} For this purpose, we consider the possibility to continuously monitor the quantum trajectories of the evolving QRC by introducing weak quantum measurements.

If information is extracted at the output layer through projective measurements of all the reservoir physical units, then most coherence is lost also erasing information of previous inputs.
As a strategy to overcome this extreme back-action effect, the QRC can be monitored by preserving part of its  fading memory either considering weak measurements  or projectively measuring only part of the reservoir nodes (indirectly measuring the rest of the reservoir)~\cite{RevModPhys.82.1155,Hatridge2013}. Both approaches are actually related~\cite{wiseman2009quantum} and here we will focus on the case of a tunable weak measurement.
Then, alternating input injection and output extraction --as in classical RC-- we propose an \textit{online} time series processing protocol based on weak measurement, as represented in panel (d) of Fig.~\ref{fig:protocols}.

For a single realization of the online protocol {(OLP)}, the monitored state  $\rho_{\mathbf{V}_k}$ (for which we drop the apex `u') evolves following a quantum trajectory characterized by the  (stochastic) measurement outcomes $\mathbf{V}_k$:
\begin{equation}
\label{eq:rhotrajectories}
    \rho_{\mathbf{V}_k}=\frac{\left(\mathcal{M}_{\mathbf{V}_k} \circ  \mathcal{L}_k\right)[\rho_{\mathbf{V}_{k-1}}]}{\textnormal{Tr}\left(\left(\mathcal{M}_{\mathbf{V}_k} \circ  \mathcal{L}_k\right)[\rho_{\mathbf{V}_{k-1}} ]\right)},
\end{equation}
where the effect of the measurements on the system is determined by the form of the superoperator $\mathcal{M}_{\mathbf{V}_k}$ (see also Eq.~(\ref{eq:ourMk})) and the state $\rho_{\mathbf{V}_k}$ is properly normalized to have unit trace.
The mixed state that accounts for all possible measurement outcomes at time step $k$, assuming an infinite ensemble of realizations, is given by:
\begin{equation}\label{eq:rhok}
\rho_k=\int_{-\infty}^{\infty}\hspace{-1em}d\mathbf{V}_k\rho_{\mathbf{V}_{k}}P(\mathbf{V}_k)
=\int_{-\infty}^{\infty}\hspace{-1em}d\mathbf{V}_k \left(\mathcal{M}_{\mathbf{V}_k} \circ  \mathcal{L}_k\right)[\rho_{k-1}],
\end{equation}
where the details of the derivation from Eq.~\eqref{eq:rhotrajectories} can be found in Appendix \ref{appendix_soleq5}.
The expectation values of operators in the OLP can be immediately generalized by replacing the unperturbed state $\rho_k^{\rm u}$ of the RSP and RWP with the monitored state $\rho_k $, accounting for the measurement back-action.
\subsection{Monitored qubit system} \label{numresA}
In order to account for the measurements during the dynamics of the system with the OLP, we introduce in Eq.~\eqref{eq:rhotrajectories} the following measurement superoperator:
\begin{equation}
\label{eq:ourMk}
   \mathcal{M}_{\mathbf{V}_k}[\rho]=\hat{\Phi}_{\mathbf{V}_k} \rho \hat{\Phi}^{\dagger}_{\mathbf{V}_k}.
\end{equation}
Given a system of $N$ qubits, single-qubit measurements in a single realization are described by the Kraus operators $\hat{\Phi}_{\mathbf{V}_k}=\hat{\Omega}_{V_k^{(0)}} \otimes\, \cdots \, \otimes \hat{\Omega}_{V_k^{(N-1)}},$ which give a set of measurement results $\mathbf{V}_{k\,l}=\{V^{(i)}_{k\,l}\}$, with $i=0,\, ...,N-1$, where the indices $k$ and $l$ label the time step and the experimental realization, respectively.

We employ indirect measurements already implemented experimentally in diverse platforms~\cite{introExpQmeasurements,SiddiqiNature,SiddiqiQtrajectories,Hatridge2013,pan2020weak,memristors_walther}. For instance, the state of a superconducting qubit coupled to a  cavity is indirectly measured by probing the cavity with a microwave signal~\cite{introExpQmeasurements,SiddiqiNature,SiddiqiQtrajectories,Hatridge2013}. In the trapped ion experiment in ~\cite{pan2020weak}, the information about the electronic states is extracted from their interaction with the vibrational motion.
For all these platforms, indirect measurements in the z direction can be modeled by means of the following operator:
\begin{equation}\label{eq:operV}
    \hat{\Omega}^{\rm z}_V=\frac{1}{\sqrt[4]{2\pi}}\left(e^{-\frac{(V-g)^2}{4}}\ket{0}\bra{0}+e^{-\frac{(V+g)^2}{4}}\ket{1}\bra{1}\right),
\end{equation}
where $g$ is the tunable strength of the measurement and $V$ is the measurement outcome, both normalized with the two Gaussians width.
Measurements in the x and y directions can be obtained by means of rotations (see Appendix~\ref{Appendix_Measurements_Formalism}).
From Eq.~\eqref{eq:operV}, the probability distribution for an outcome $V$, when we measure a qubit in the reduced state $\omega$, $P_{\rm z}(V)= \text{Tr}(\hat{\Omega}^{\rm z\dagger}_V\hat{\Omega}_V^{\rm z}\omega)$, is:
\begin{equation}\label{eq:P_V}
P_{\rm z}(V)=
\omega_{00}\frac{1}{\sqrt{2\pi}}e^{-\frac{(V-g)^2}{2}}+(1-\omega_{00})\frac{1}{\sqrt{2\pi}}e^{-\frac{(V+g)^2}{2}},
\end{equation}
where $\omega_{00}=\langle 0 |\omega |0 \rangle$, {and whose} outcomes are sampled from a weighted sum of two Gaussian distributions (at distance $2g$) depending on the state before the measurement. Therefore, in the limit of large $g$, the projective case is approached and most outcomes $V$  can be unambiguously assigned to one of the two distributions. However, if $g$ is small, the measurement is weak and the two distributions largely overlap making the measurement less informative. 
With the outcomes obtained from indirect measurements, we can compute the expectation values of the observables. For instance, when we measure a single qubit with state $\omega$ in the z direction, we have
$ \langle V \rangle_{\omega}^{\rm z}=\int_{-\infty}^{\infty}V \, P_{\rm z}(V)\, dV=g\langle \hat{\sigma}^{\rm z} \rangle_{\omega}$.
The expectation values of two-qubit observables are computed in a similar way, as shown in Appendix~\ref{appendix_twoqubit_measurements}.

By using Eq.~\eqref{eq:operV} in Eq.~\eqref{eq:rhok} (see Appendix~\ref{appendix_soleq5}), we can evaluate the integrals and leave a clean formulation of the iterative map
\begin{equation}
\label{eqrhoidealwithg}
\rho_k=\int_{-\infty}^{\infty}\hspace{-1em}d\mathbf{V}_k \hat{\Phi}_{\mathbf{V}_k} \mathcal{L}_k[\rho_{k-1}] \hat{\Phi}^{\dagger}_{\mathbf{V}_k}=M\odot \mathcal{L}_k[\rho_{k-1}],
\end{equation}
where $\odot$ represents the element-wise product and $M$ is defined as
\begin{equation}\label{eq:M}
   M\equiv \tilde{M}^{\otimes N}, \quad \tilde{M}=\begin{pmatrix}
1&e^{-\frac{g^2}{2}}\\
e^{-\frac{g^2}{2}}&1
\end{pmatrix}.
\end{equation}
This allows us to quantify the increasing decoherence introduced by sharper measurements, while for very small $g$ the state is weakly perturbed.

An interesting question is about the possible effect of measurement as a further non-linearity source for QRC.
In order to more clearly address this point, let us specialize to the framework of QRC introduced in Ref.~\cite{Nakajima2017}, which will be also used in the following sections. The reservoir consists of a  qubit network unitarily evolving under the disordered transverse-field Ising Hamiltonian~\cite{Nakajima2017,martinez2021dynamical}, while the input is injected into the system by rewriting a node state $\rho_k^{\textnormal{in}}$. The system parameters are chosen in such a way that the reservoir is found in the ergodic dynamical phase, which has been shown to guarantee optimal QRC performances~\cite{martinez2021dynamical}.
Different realizations of the reservoir produce similar results.
Full details are given in Appendices~\ref{MethodsSec} and \ref{networkrealizations}. 
By using Eqs.~(\ref{eqrhoidealwithg}) and (\ref{eq:M}), we find that  the explicit dependence on the input and on $g$ of the components of $\rho_k$  is translated to the observables as (see Appendix~\ref{appendix_soleq5}):
\begin{equation}
\label{eqobsvsskandg}
    \langle \hat{\mathcal{O}} \rangle_{\rho_k}^{\infty}=\sum_{n=1}^{N}e^{-\frac{N-n}{2}g^2}\left(A_{n}+B_{n}s_{k}+C_{n}r_k\right),
\end{equation}
where $s_k$ is the $k$th input injected, $r_k\equiv\sqrt{s_k(1-s_k)}$, and the matrices $A_n$, $B_n$, and $C_n$ depend on the state $\rho_{k-1}$, as it was found in the absence of measurement in Ref.~\cite{Mujal2021nonlinearities}.
Iterating these relations, it is then found that no further nonlinearities beyond the polynomials of $r_k$ and $s_k$ with different $k$ are introduced in the measurement process. The exponential factors of  Eq.~(\ref{eqobsvsskandg}) effectively determine the amount of measurement-induced back-action introduced in the system, being all equal to 1 in the unperturbed case.
\subsection{Measurement accuracy} \label{sec:accu}
In the previous discussion, we have anticipated two main experimental limiting factors to the detection of the  expected values of the system observables: (i) the finite number of stochastic measurements in the ensemble, affecting all the protocols of Fig.~\ref{fig:protocols}, and (ii) the measurement weakness, which preserves better the state but limits the amount of information that can be extracted and is particularly relevant for the OLP.

Within our measurement framework, for single-qubit observables,  these two sources of statistical uncertainty combine together to give an error up to (see Appendix~\ref{Appendix_Measurements_Formalism}):

\begin{equation}
\label{eqstatisticalsinglequbits}
\bar{s}_{\langle\hat{\sigma}\rangle}=\sqrt{\frac{g^2+1}{g^2 N_{\rm meas}}},
\end{equation}
while for two-qubit observables the statistical uncertainty reads:
\begin{equation}
\label{eqstatisticaltwoqubits}
\bar{s}_{\langle\hat{\sigma}\otimes\hat{\sigma}\rangle}=\sqrt{\frac{g^4+2g^2+1}{g^4 N_{\rm meas}}}.
\end{equation}
The fact that more information is extracted with sharper measurements ($g\gg 1$) is reflected in a reduction in the statistical uncertainty in the observables when increasing $g$, for a fixed number of measurements. For the RSP and the RWP, this justifies the use of strong measurements, approximately projective (e.g. $g=10$). However, for the OLP a careful assessment is needed, taking into account that measurement back-action in general leads to deviations from the unperturbed dynamics, erasing information. Equations \eqref{eqstatisticalsinglequbits}-\eqref{eqstatisticaltwoqubits} also tell us that the  degree of precision depends on the kind of observable chosen. Indeed, taking a weak value $g\ll1$, the same uncertainty obtained with $N_{\rm meas}\sim g^{-2}$ for single-qubit observables would require    $N_{\rm meas}\sim g^{-4}$ in the case of two-qubit  observables.

\begin{figure}[t!]
    \centering
    \includegraphics[width=\columnwidth]{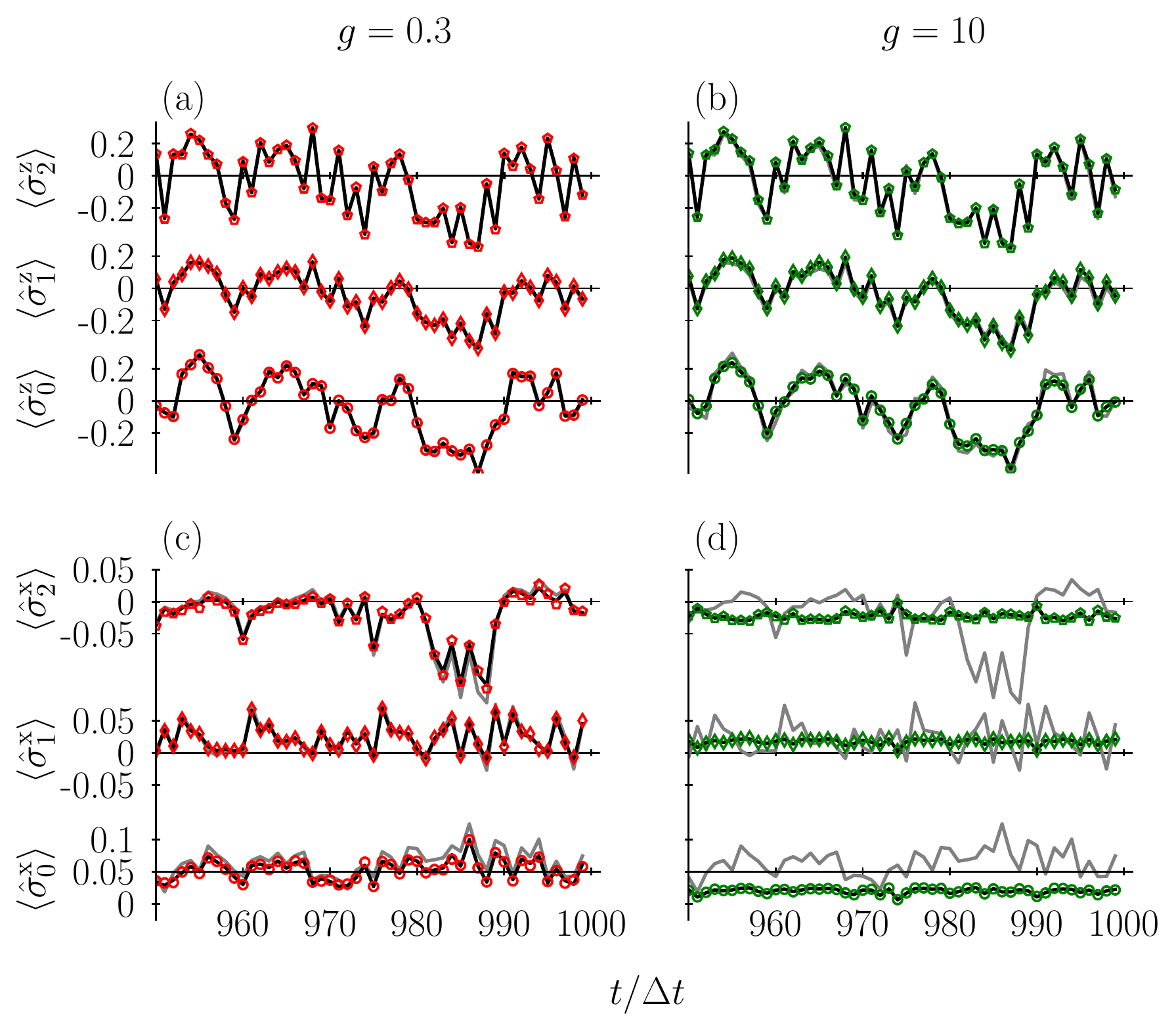}    \caption{Single-qubit observables of the first three qubits of the system with $N=6$ in the (a)-(b) z and (c)-(d) x directions. Solid lines correspond to the ideal values ($N_{\rm meas}\rightarrow \infty$) with (black) the effect of measurements on the system (OLP) and (grey) the unperturbed case (RSP and RWP), respectively. In symbols, the values obtained by numerically simulating the measurement process with $N_{\rm meas}=1.5 \cdot 10^6$ in the OLP. Panels (a), (c) correspond to a weak measurement strength, $g=0.3$; whereas in panels (b), (d) strong measurements, $g=10$, are performed. The input is encoded into the state of qubit 0 (see Appendix~\ref{MethodsSec}).}
    \label{fig:singlequbitobservables}
\end{figure}

In order to appreciate the way back-action and the two sources of statistical uncertainty (i) and (ii) affect the OLP, we plot in 
Fig.~\ref{fig:singlequbitobservables} the dynamics of some observables and compare them with the ideal quantities.
The finite number of measurements is responsible for small deviations (not visible in the plot), namely, $\bar{s}_{\langle\hat{\sigma}\rangle}\approx 2 \cdot 10^{-3}$ in panels (a) and (c), and $\bar{s}_{\langle\hat{\sigma}\rangle}\approx 8 \cdot 10^{-4}$ in (b) and (d). Actually, as we will show below, these seemingly small fluctuations have an important impact for machine learning purposes. The most apparent discrepancies are due to measurement back-action and are shown in Fig.~\ref{fig:singlequbitobservables}(d), between the (OLP) measured and the (RSP and RWP) unperturbed values.
Back-action impact is rooted in the reservoir state features and finally in its operation regime:
In the z-direction, in panels (a) and (b), this effect is less important because the qubits are mostly aligned along z due to the reservoir parameter choice \cite{martinez2021dynamical}.
If the measurements are done in a different direction, i.e. to gather $\langle \hat{\sigma}^{\rm x}_i \rangle$ and are sharp (Fig.~\ref{fig:singlequbitobservables}(d)), they introduce strong decoherence leading to the strong deviation with respect to the unperturbed state.
Only with sufficiently weak measurements, as in Fig.~\ref{fig:singlequbitobservables}(c), the (unperturbed) dynamics is qualitatively and approximately preserved with the OLP.  A similar discussion also applies to more general observables, like two-qubit observables (see Appendix~\ref{appendix_observables}).
These analytical and numerical findings hint at the fact that a weak measurement is the only viable option for the OLP.

An important consequence of Eq.~(\ref{eq:M}) is that, for $g\ll 1$, the ideal state (monitoring an infinite ensemble) is weakly perturbed, as $M$ approaches the unity matrix with all components 1.
Intuitively, the disturbance introduced by such measurement is so weak that almost no information is lost during the online processing.
With finite measurement  ensembles $N_{\rm meas}$, the error  associated to this unsharp measurement is larger than in the unperturbed state (e.g. $\bar{s}_{\langle\hat{\sigma}\rangle}$ in Eq.~(\ref{eqstatisticalsinglequbits}) is larger than $1/\sqrt{N_{\rm meas}}$). Interestingly, in the ideal case (infinite resources)  information would still be extracted with a limit vanishing error $\bar{s}_{\langle\hat{\sigma}\rangle}$. Building on this analysis, the estimation of observables experimentally accessible with finite measurements can be numerically emulated  adding the proper Gaussian noise to the ideal ones, $\langle \hat{\mathcal{O}}\rangle_{\rho_k}^{\infty}$. The comparison between either performing numerically an ensemble over $N_{\rm meas}$ independent trajectories or adding a Gaussian noise to an ideal infinite ensemble (with standard deviation given by either Eq.~\eqref{eqstatisticalsinglequbits} or \eqref{eqstatisticaltwoqubits}) will also be verified below.
\subsection{Short-term memory capacity}\label{sec:STM}
In the previous sections, we have analytically assessed measurement effects and  we are now applying this framework to QRC assessing its memory (this section) and forecasting (next section) capabilities.
In order to identify the accuracy limitations and back-action effects of  measurements, we consider the short-term memory (STM) task. This allows us to establish the reservoir efficiency in storing past information, that is, the fading memory crucial for effective reservoir computing.
Given the input sequence $\{s_k\}$ of random numbers generated from a uniform distribution, the task is to reproduce the target values
\begin{equation}
y_k=s_{k-\tau},
\end{equation}
where $\tau$ is the delay. The predictions are obtained with different measured observables (for instance $\langle\hat{\sigma}^{\rm x}_i\rangle$, or $\langle\hat{\sigma}^{\rm z}_i\hat{\sigma}^{\rm z}_j\rangle$) for all qubits $i,j=1,N$ and comparing RSP, RWP and OLP. 
\begin{figure}[t!]
    \centering
       \includegraphics[width=\columnwidth]{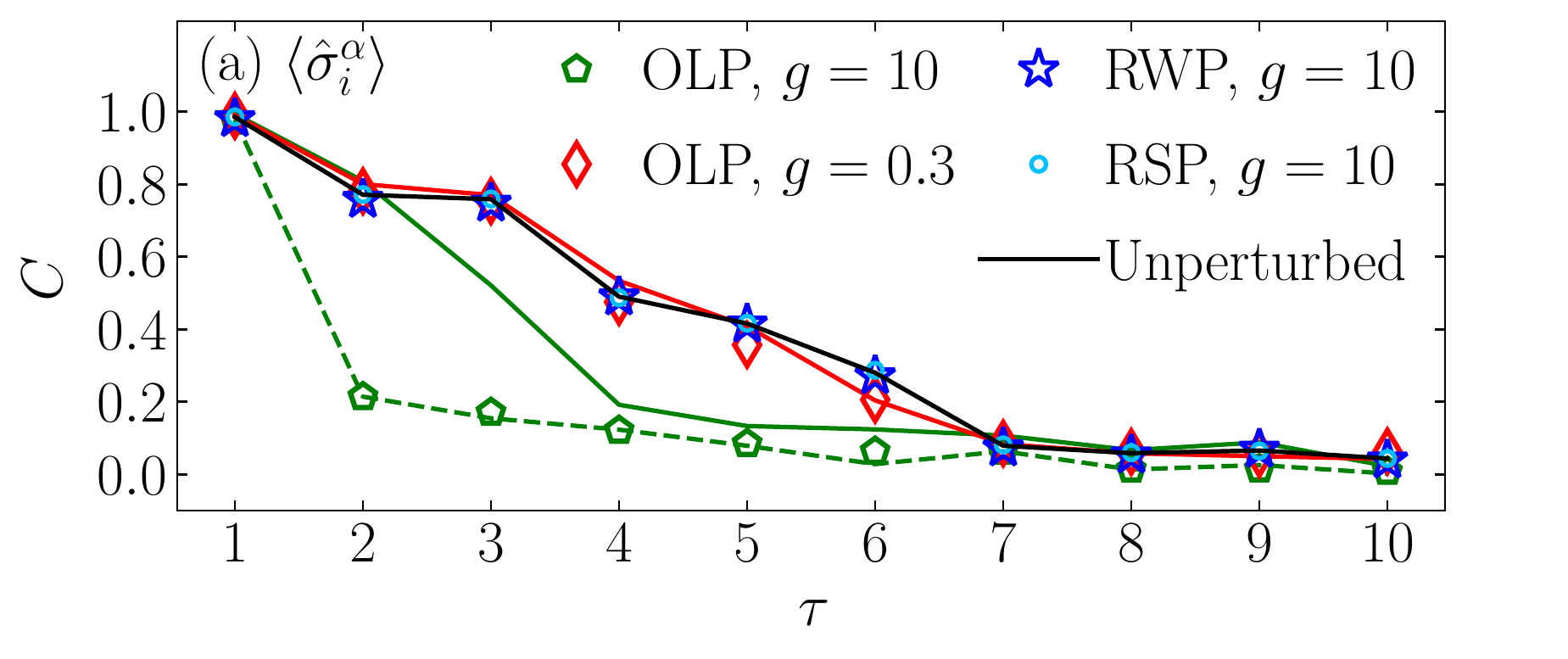}    \\
       \includegraphics[width=\columnwidth]{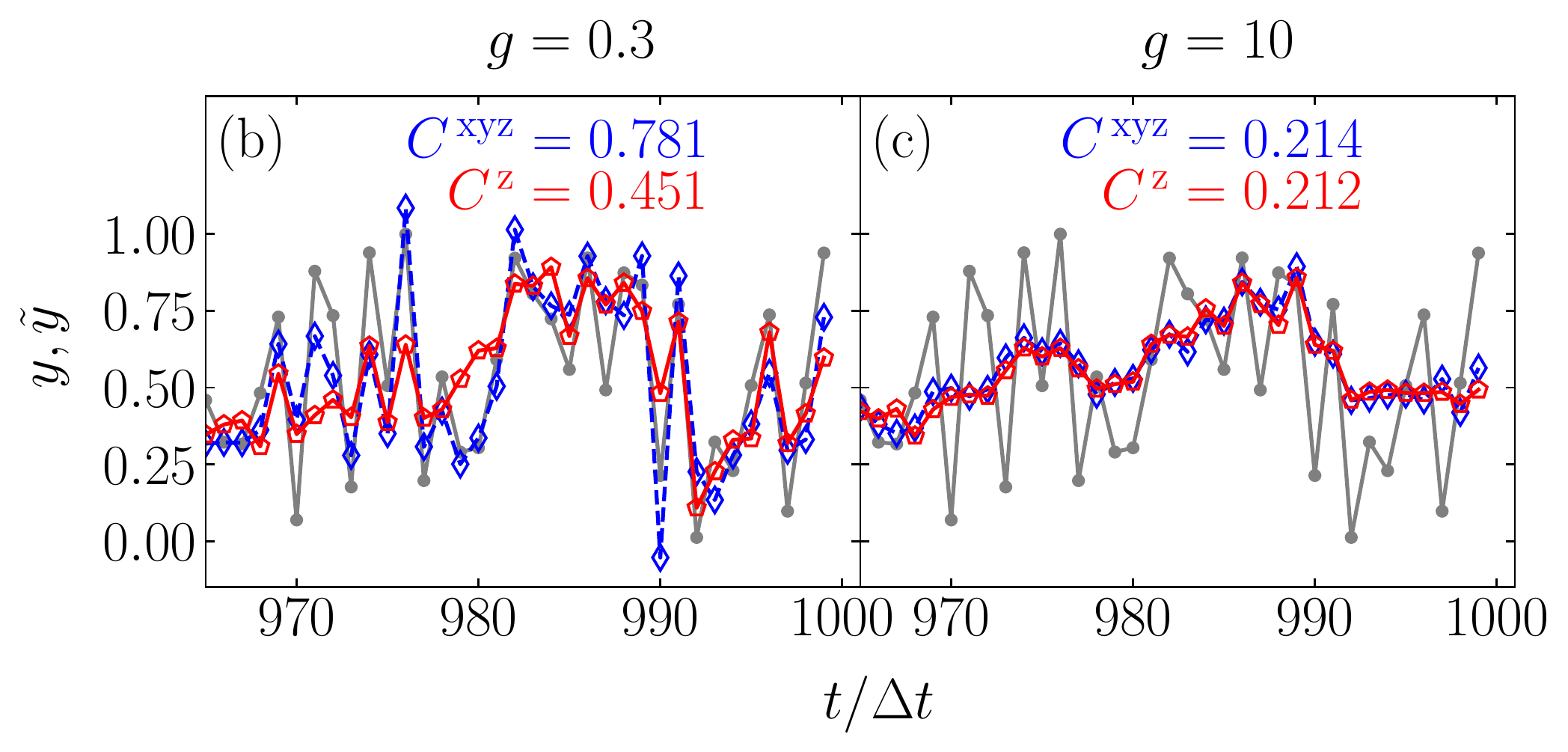}
    \caption{(a) STM capacity from single-qubit observables depending on the delay $\tau$ obtained with the different protocols and $N_{\rm meas}=1.5\cdot 10^6$ measurements (symbols). The ideal capacities ($N_{\rm meas}\rightarrow \infty$), are represented with a solid line in the corresponding color; the unperturbed situation is in black. The dashed green line corresponds to estimated values obtained from the observables in the limit $N_{\rm meas}\rightarrow \infty$ but adding a Gaussian noise. Indices of label $\langle \hat{\sigma}^{\alpha}_i  \rangle$ refer to $i=0,\, ...,N=6$ and $\alpha=$ x, y, z. In (b) and (c), target values (grey filled circles) are shown with predictions (colored unfilled symbols) for the STM task at $\tau=2$ obtained with the OLP and weak measurements (b), and strong measurements (c), both with single-qubit observables. All data shown correspond to the test data set.}
    \label{fig:targetpred_capacities_vs_tau}
\end{figure}
As usual, the performance is quantified comparing predictions, $\tilde{\bm y}=\{\tilde{y}_k\}$, and target values, ${\bm y}=\{y_k\}$, through the normalized correlation, namely the capacity  $C(\tau)\in[0,1]$. Perfect STM at delay $\tau$ corresponds to $C(\tau)=1$. More details in Appendix~\ref{MethodsSec}.

The STM capacity for increasing delays is shown in  Fig.~\ref{fig:targetpred_capacities_vs_tau}(a), for single-qubit observables in the output layer. The performance when neglecting any perturbation introduced by measurement (black line) progressively decays, showing the ability of the reservoir to store past information up to about six previous steps (consistently with the network size $N=6$ \cite{martinez2020information}). This STM capacity can be achieved with the RWP and the RSP with high accuracy for $N_{\rm meas}=1.5\cdot 10^6$. We remark that values of $N_{\rm meas}$ ranging from $10^4$ to $10^6$ are in accordance with experiments of quantum computation \cite{arute2019quantum} and quantum machine learning \cite{havlivcek2019supervised}. When continuously monitoring the system in the OLP, the effect of measurement back-action becomes critical and for sharp measurement ($g=10$) the performance is significantly hindered.
Remarkably, this limitation is overcome with weak measurements and the STM capacity achieved for $g=0.3$  is  approaching the unperturbed case. Therefore, weak measurements  preserve short memory performing this \textit{online} temporal task, sequentially injecting new inputs and continuously extracting information. The explanation of the hindered performance for strong measurement ($g=10$) is shown to be due to both back-action and statistical noise, {with $C$ decaying strongly for $\tau=2$}. Assuming infinite accuracy (ideal case of infinite $N_{\rm meas}$, green continuous line in Fig.~\ref{fig:targetpred_capacities_vs_tau}(a)), the capacity {instead reaches larger delays ($\tau=4$). Therefore}, the ideal OLP with $g=10$ does not achieve the memory obtained with the other protocols due to back-action:  continuous sharp measurements erase reservoir information encoded in quantum coherences and correlations (Fig.~\ref{fig:singlequbitobservables}(d)).

The input sequence to the QRC and the ability to reproduce past inputs in the OLP are shown in more detail in  Fig.~\ref{fig:targetpred_capacities_vs_tau}(b) and (c), for the STM capacity at delay $\tau=2$.  For weak measurement (b), all qubits observables ($\hat{\sigma}^{\rm x},\hat{\sigma}^{\rm y},$ and specially $\hat{\sigma}^{\rm z}$) contribute to the capacity $C(\tau=2)$, while sharp measurements (large $g$) hinder the ability of the output to reproduce the target.
On the one hand, Fig.~\ref{fig:targetpred_capacities_vs_tau}(b) also tells us that, despite the state is mostly aligned in the z direction  \cite{martinez2021dynamical}, the reservoir is far from approaching the classical regime, where quantum coherences would be negligible. Indeed, significant contributions to the capacity come from all the qubit directions (more details are given in Appendix~\ref{appendix_nonclassical}).
Intuitively, even though  such coherences are small with respect to the populations (see also Fig.~\ref{fig:singlequbitobservables}),  they still provide significant richness to the reservoir output (being larger than the statistical uncertainty of the measurement process). On the other hand, we see in  Fig.~\ref{fig:targetpred_capacities_vs_tau}(c)  that if such coherences
are further hindered by stronger back-action (falling below the statistical noise, then the capacity reached with $\hat{\sigma}^{\rm z}$ ($C=0.21$) does not improve when also considering $\hat{\sigma}^{\rm x}$ and $\hat{\sigma}^{\rm y}$.
\begin{figure}[t!]
    \centering
    \includegraphics[width=\columnwidth]{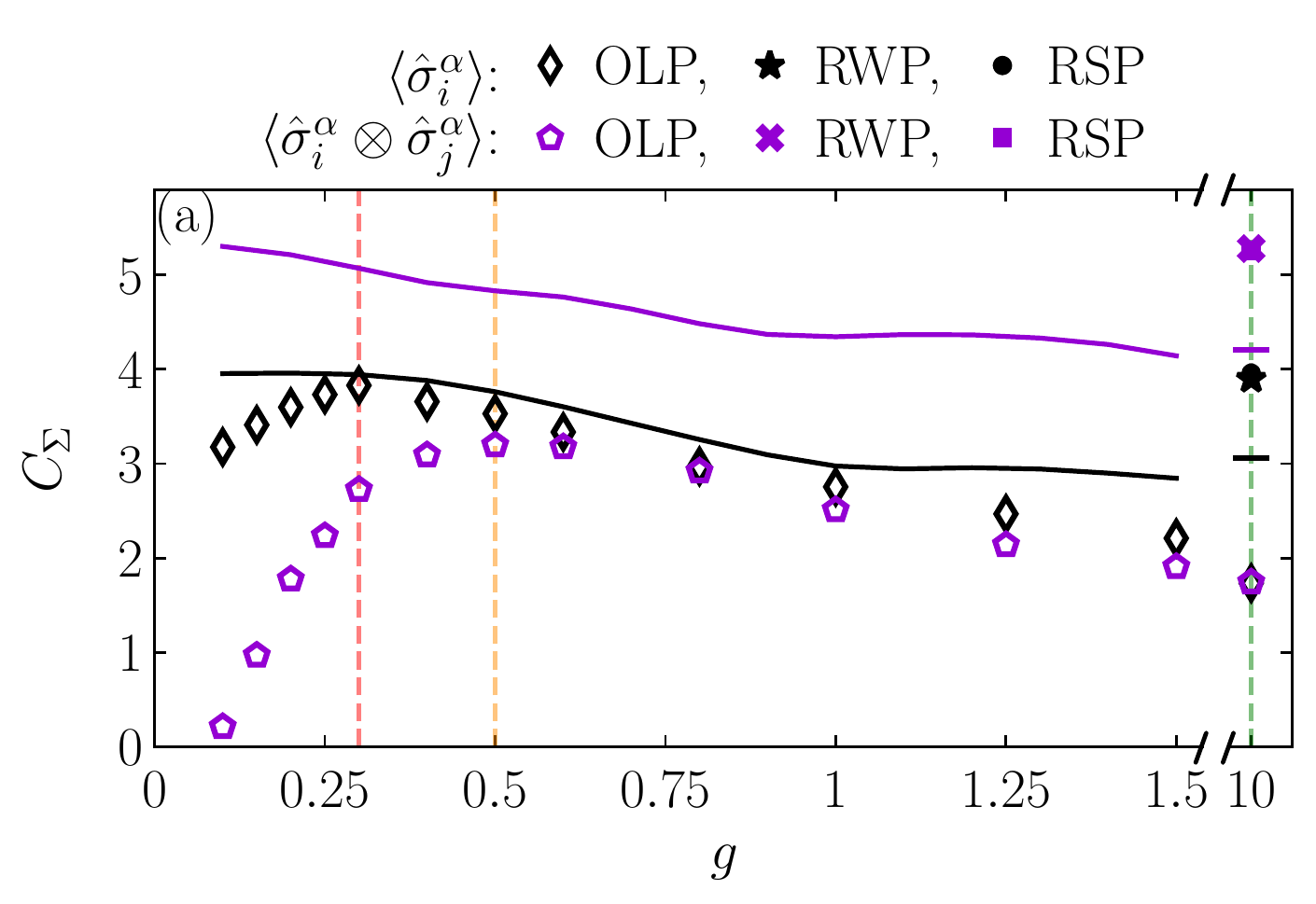}
    
    \vspace{0.15cm}
    
    \includegraphics[width=\columnwidth]{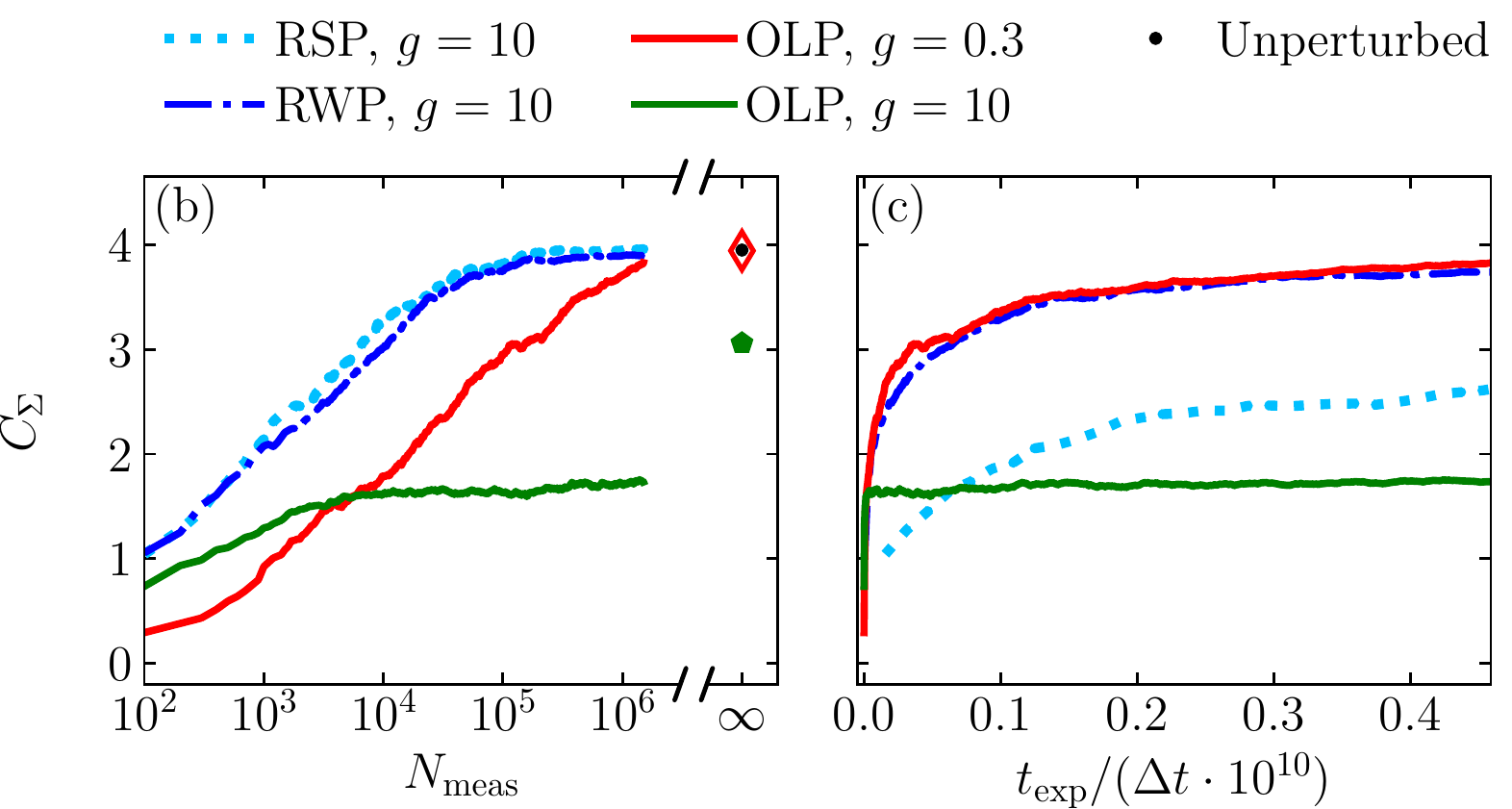}
    \caption{(a) STM sum capacity depending on $g$ for different kind of observables. For the OLP, solid lines represent the values when $N_{\rm meas}\rightarrow \infty$ associated to the unfilled symbols in the same color for the $N_{\rm meas}=1.5\cdot 10^6$ case. Vertical dashed lines represent the values of $g=0.3$ (red), $g=0.5$ (orange), and $g=10$ (green). The capacities of the RWP and the RSP are nearly identical. This figure has also been reproduced for 100 realizations of the random couplings of the Hamiltonian, see Fig.~\ref{fig:averages}. In (b), $C_{\Sigma}$ is plotted as a function of the number of measurements; in (c), depending on the experimental time required. Single-quibt observables, $\langle \sigma^{\alpha}_{i}\rangle$, are used in (b) and (c), with $\alpha=$ x, y, z and $i=1,\, ..., N=6$. All data shown correspond to the test set.}
    \label{fig:sumcapacityvsnmeasandtexp}
\end{figure}

The overall performance of the STM task is quantified in Fig.~\ref{fig:sumcapacityvsnmeasandtexp} by the sum capacity {accounting for the total memory at all  delays (equivalent to truncating at 10 as $C(\tau > 10)\sim 0$)}
\begin{equation}
C_{\Sigma}\equiv \sum_{\tau} C(\tau),
\end{equation}
as a function of $g$, considering both single-qubit and two-qubit observables.
In the infinite-measurements limit, the OLP (solid lines) in Fig.~\ref{fig:sumcapacityvsnmeasandtexp}(a) displays maximum memory $C_{\Sigma}$ for $g\rightarrow 0^+$, when the system becomes effectively unperturbed (no back-action effect, see Eq.~(\ref{eq:operV})). However, in realistic implementations with a finite number of measurements,  the statistical uncertainty diverges in this extremely weak (non-informative) measurement limit as shown in Eq.~\eqref{eqstatisticalsinglequbits}. Then, the optimum sum capacity is achieved at a larger $g$ value, depending on $N_{\rm meas}$, where the output expectation values are more accurate but at the same time the back-action is not too strong.
Indeed the larger $N_{\rm meas}$, the weaker the measurement that maximizes the capacity.
With $N_{\rm meas}=1.5\cdot 10^6$, the optimal measurement strength is found around $g=0.3$ for one-qubit observables (black diamonds).

Two-qubit correlations can also be exploited for QRC, and, in the ideal limit (solid purple line in Fig.~\ref{fig:sumcapacityvsnmeasandtexp}(a)) they reach a larger memory capacity than single-qubit ones, as expected for a larger number of observables. However, these correlations are smaller and subject to a  more significant statistical uncertainty. Therefore, with a finite number of measurements, the performance is actually worse than for single qubit observables (symbol lines). 
Of course, we also see that in this case there is further room for improvement by increasing $N_{\rm meas}$.
This is a key point in view of experimental implementations as, beyond the performance achievable in QRC when considering quantum measurements, it is critical to quantify the amount of needed resources. 
In Fig.~\ref{fig:sumcapacityvsnmeasandtexp}(b), we represent the sum capacity as a function of the number of measurements, showing that with the OLP limited to weak measurements (red line), as well as with the RSP and the RWP, the sum capacity increases as soon as the statistical uncertainties become negligible and until the corresponding upper bound is approached. Indeed, while for ensemble sizes $N_{\rm meas} \sim 10^4$, reset protocols perform better, all the RSP, the RWP and the OLP with $g=0.3$ reach ideal capacities for $N_{\rm meas} \sim 10^6$.
In contrast, the use of strong measurements with the OLP leads to a slow convergence of the capacity (green line in Fig.~\ref{fig:sumcapacityvsnmeasandtexp}(b)), which seems to saturate to a lower value than the one corresponding to the $N_{\rm meas}\rightarrow \infty$ limit.

In order to analyze the real amount of resources demanded by each protocol, it is meaningful to quantify the  time needed for each experimental realization (see Appendix~\ref{appendix_gvaluesandtwo}, Eqs.~\eqref{texp_restartingAPPENDIX}--\eqref{texp_onlineAPPENDIX}) and not only the ensemble size $N_{\rm meas}$. 
The required experimental time for each measurement protocol is shown in Fig.~\ref{fig:sumcapacityvsnmeasandtexp}(c) for the case of a single-observable output layer.
As expected, the RSP is not efficient, while both the OLP and the RWP achieve a good performance with the same resources, if the measurement is weak in the online approach (OLP with $g=0.3$). Indeed it is interesting to compare the OLP and the RWP by finding the value of $g$ such that $t^{\rm OLP}_{\textnormal{exp}} \leq t^{\rm RWP}_{\textnormal{exp}}$ while also imposing  the same statistical uncertainty in both cases. This leads to finding a criterion to determine the minimum measurement strength for online processing to be advantageous with respect to rewinding. The relevant quantity to this aim is the number of washout steps $N_{\rm wo}$, (Appendix~\ref{appendix_gvaluesandtwo}) that, as shown in Sec.~\ref{sec:prot}, defines the finite memory of the reservoir.
For single-qubit observables we find 
\begin{equation}\label{eq:gthr_1}
g \geq \sqrt{\frac{1}{N_{\textnormal{wo}}-1}}.
\end{equation}
This is a necessary condition to be successful in online processing (with less resources than the RWP), while it will be sufficient if the corresponding measurement is weak enough to have negligible back-action effects in the OLP.
For instance, if the washout steps amount to $N_{\textnormal{wo}}=20$, the measurement strength in Eq. \eqref{eq:gthr_1} is $g \gtrsim 0.23$ producing negligible back-action. Indeed the red line of Fig.~\ref{fig:sumcapacityvsnmeasandtexp}(c) of best capacity corresponds to $g=0.3$.
The bound in $g$ for efficient OLP also depends on the kind of observables considered.
The analogous condition for two-qubit observables is more restrictive, i.e. a stronger measure is needed, as described in Appendix~\ref{appendix_gvaluesandtwo} and further discussed in Appendix~\ref{appendix_observables}.
\subsection{Chaotic time series forecasting}\label{sec:santafe}
\begin{figure}[t!]
    \centering
    \includegraphics[width=\columnwidth]{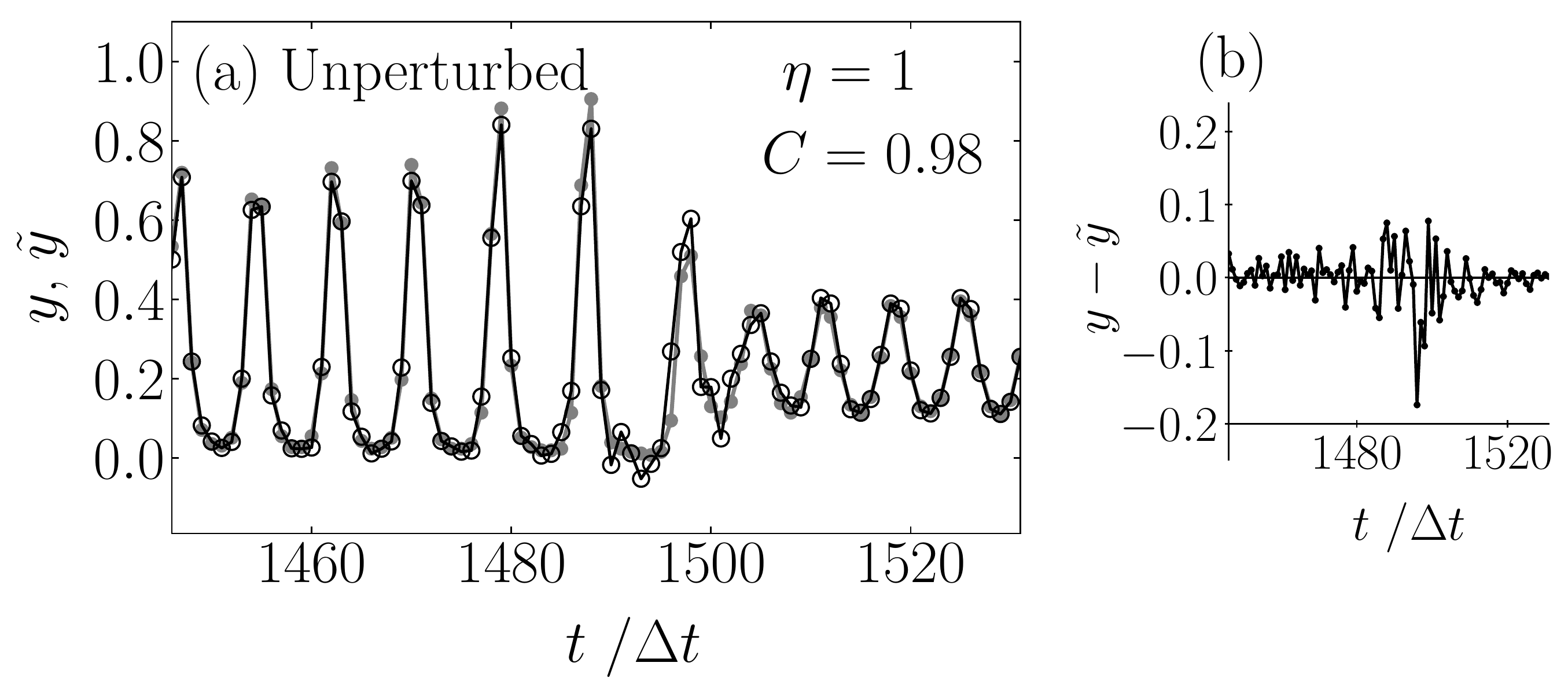}
        
    \vspace{0.2cm}
    
    \includegraphics[width=0.92\columnwidth]{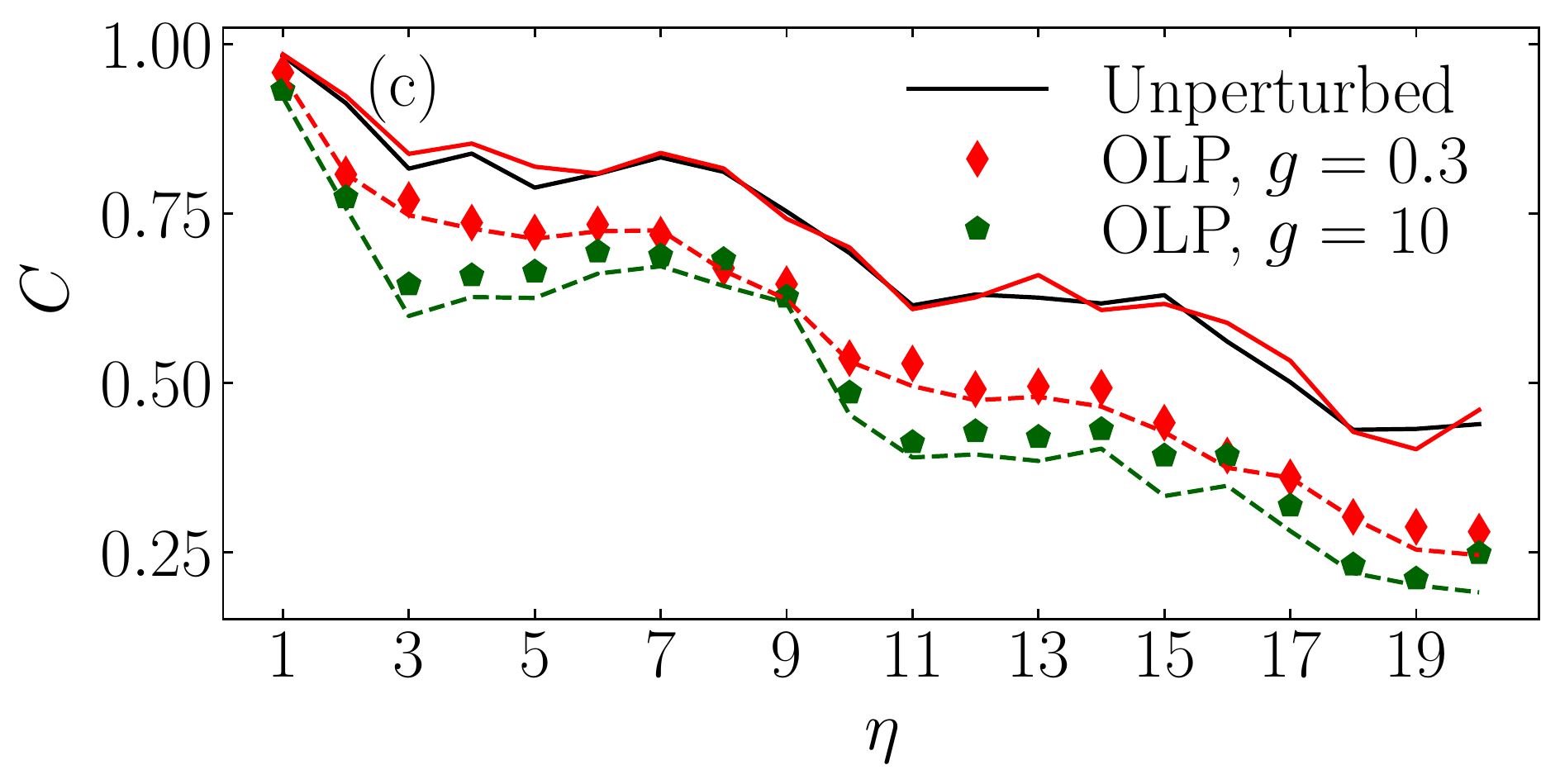}
    \caption{(a) One-step forward predictions and (b) prediction errors on the Santa Fe time series for a part of the test data when the observables of the unperturbed quantum reservoir are employed. In panel (a), target values  correspond to grey filled circles and predictions to black unfilled circles. (c) Capacity of predictions, $C$, depending on the number of time steps forward $\eta$. In panel (c), the OLP with $g=0.3$ and $g=10$ is used, respectively, to estimate the observables with $N_{\rm meas}=1.5\cdot 10^6$ measurements at each time step. The limit case for $g=0.3$ with $N_{\rm meas}\rightarrow \infty$ is represented with a solid red line; the unperturbed situation is in black. The dashed red and green lines correspond to estimated values obtained in the limit $N_{\rm meas}\rightarrow \infty$ but adding a Gaussian noise to the observables. The values of the capacities correspond to the test data.}
    \label{fig:santfe}
\end{figure}
After the analysis of the STM task, we quantify the capacity of the QRC system to make forward predictions on the Santa Fe laser time series, corresponding to a NH$_3$ laser in a chaotic dynamical state~\cite{PhysRevA.40.6354}. 
This benchmark dataset originates from a time-series forecasting competition held at the Santa Fe Institute in the 1990s~\cite{298828} and it has been used extensively ever since.
Formally, our target is written as
\begin{equation}
    y_k=s_{k+\eta},
\end{equation}
where $\eta$ labels the number of steps forward to make the predictions. The original values of the Santa Fe series were shifted and normalized such that $\{s_k\}\in [0,1]$ (see Appendix~\ref{MethodsSec}).

In Figs.~\ref{fig:santfe}(a) and (b), we illustrate the performance of the QRC for one-step forward predictions with the unperturbed system. By considering together both single-qubit and two-qubit observables, a remarkable capacity of $C=0.98$ is reached. A similar accuracy can be obtained with the RSP or the RWP.
The prediction of the Santa Fe time series is particularly challenging when there is a jump from large to low amplitude oscillations, see $t/\Delta t \sim 1500$ in Fig.~\ref{fig:santfe}(a).
Accordingly, the most appreciable differences in Fig.~\ref{fig:santfe}(b) between the target and the predicted values are found when there are sudden changes in the amplitude of the signal, which are nevertheless mostly captured by the predicted values.

The quality of further future predictions ($\eta\geq1$) is shown in Fig.~\ref{fig:santfe}(c), where we have concentrated on the study of the results obtained with the OLP in comparison to the unperturbed situation (ideally retrieved by the RSP and the RWP).
The prediction task becomes more complicated as the prediction distance $\eta$ increases.
Accordingly, the goodness of the predictions decreases with increasing $\eta$. 
As an exception, there are local maxima in the capacity due to the quasi-periodicity in the Santa Fe laser time series, i.e. at $\eta\approx7$ and $\eta\approx14$.
As shown in the previous section for the STM task, the perturbative nature of a sharp quantum measurement (e.g. $g=10$) erases part of the QRC memory on the input history.
The results presented in Fig.~\ref{fig:santfe}(c) illustrate that such a memory loss also impacts the forecasting capabilities of the QRC, with the prediction capacity for 
$g=10$ falling below the one for $g=0.3$.
Interestingly, in the limit of infinite measurements, the capacity for the weak measurement  matches the one of the unperturbed system and there is no significant back-action effect on future predictions. The considerably good capacity obtained with the simulated experimental data with weak measurements in Fig.~\ref{fig:santfe}(c) (red symbols) could be improved with more measurements to eventually approach the upper bound (solid red line).
\section{\label{sec:conclusions} Conclusions}

The potential advantage in processing temporal data sequences with quantum systems in reservoir computing is rooted in the large processing capability opened by their Hilbert space \cite{Nakajima2017,mujal2021opportunities}. 
Here, we provide the first evidence that this advantage can still be achieved beyond ideal situations, when including the effect of quantum measurement, paving the way to experimental demonstrations of sequential data processing both in memory and forecasting tasks. We have analyzed protocols where after each detection the reservoir dynamics is repeated for the whole past input sequence \cite{suzuki2022natural,chen2020temporal} (named restarting protocol, RSP) or for the last part of it, exploiting the fading memory of the reservoir (rewinding protocol, RWP).
An alternative proposal is based on weak measurements, allowing for online data processing (OLP) without repeating dynamical steps or storing any inputs externally.

For time series processing, the impact on the task accuracy (i) of statistical noise (ensemble size),  (ii) of measurement weakness, and (iii) of the back-action for continuous monitoring, need to be taken into account to design a quantum reservoir computer with a good compromise between performance and resource cost. 
Optimum performance has been shown to be reachable beyond ideal assumptions both in the RSP and RWP, with projective measurements and buffering input data, 
as well as in the OLP with (weak) monitoring. The distinctive feature of the OLP is the online operation, without storing any input, either classical or quantum. This is particularly advantageous in platforms where system ensembles can be measured at the output layer, at each input injection,
as for atomic/molecular ensembles \cite{Negoro2021} or in multimode (pulsed) photonics \cite{JGarciaBeni2022inpreparation}, where fully online time series processing with weak measurements could be realized. 

Online weak measurements enable us to partially preserve quantum coherences in the OLP. They can be implemented for different reservoirs, with a prominent example of superconducting qubits and trapped ions, where the strength of the measurement can be tuned through the interaction time between system and measurement apparatus \cite{SiddiqiQtrajectories,pan2020weak}, or through the reflectivity in the photonic memristor set-up in Ref.~\cite{memristors_walther}.
Alternative strategies for continuous monitoring time series could be considered depending on the choice of the physical reservoir, as for instance detection based on quantum jumps, while quantum non-demolition measurements could be used but are typically limited in the number of accessed observables. In general quantum trajectories have been successfully implemented in a variety of contexts \cite{SiddiqiNature,Minev2019,PhysRevLett.114.223601,PhysRevLett.57.1699,Gleyzes2007} laying the grounds for monitored QRC experiments, as recently proposed for non-temporal tasks \cite{khan2021physical}.

A further crucial aspect in view of experimental implementations of high performance QRC is the efficiency in terms of experimental resources. 
The potential advantage of QRC, given by the exponential size of the Hilbert state, requires the use of higher-order terms $\langle\hat{\sigma}^{\alpha_i}_i \otimes \cdots \otimes \hat{\sigma}^{\alpha_j}_j \otimes \hat{\sigma}^{\alpha_l}_l\rangle$ that are typically of small magnitude and actually necessitate more resources to be accurately determined. Depending on the reservoir fading memory and the performing task, we find that the OLP with measurements of the proper strength could exceed the performance of the RWP being efficient even for higher-order moments (ultimately exploiting the full size of the Hilbert space).

Quantum models suited for QRC as the one studied here are realized in state-of-the-art experiments, such as in superconducting qubits~\cite{PhysRevLett.120.050507} and trapped ions~\cite{Smith2016,Zhang2017}.
Besides our particular scheme, the realistic measurement framework analyzed here establishes the advantage of quantum reservoirs beyond ideal scenarios, and is expected to have an impact on the experimental implementations of previous~\cite{Nakajima2017,tran2021learning,chen2020temporal,martinez2021dynamical,nokkala2021gaussian,suzuki2022natural} and future proposals of time-series processing with quantum systems, including implementations of quantum recurrent neural networks.
\begin{acknowledgments}
We acknowledge the Spanish State Research Agency, through the Severo Ochoa and Mar\'ia de Maeztu Program for Centers and Units of Excellence in R\&D (MDM-2017-0711) and through the  QUARESC project (PID2019-109094GB-C21 and -C22/ AEI / 10.13039/501100011033).
 We also acknowledge funding by CAIB through the QUAREC project (PRD2018/47).
GLG is funded by the Spanish  Ministerio de Educaci\'on y Formaci\'on Profesional/Ministerio de Universidades   and  co-funded by the University of the Balearic Islands through the Beatriz Galindo program  (BG20/00085). The CSIC Interdisciplinary Thematic Platform (PTI) on Quantum Technologies in Spain is also acknowledged.
\end{acknowledgments}
\appendix
\section{Equations of the ideal case with back-action}\label{appendix_soleq5}
Here, we will derive the equations for the ideal case of sampling an infinite number of measurements, but introducing the effect of the back-action. In particular, Eqs.~\eqref{eq:rhok}, \eqref{eqrhoidealwithg}, \eqref{eq:M} and \eqref{eqobsvsskandg} are obtained step by step. We start by Eq.~\eqref{eq:rhok}:
\begin{align}\label{eq:E1}
&\rho_k=\int_{-\infty}^{\infty}\hspace{-1em}d\mathbf{V}_k \, \rho_{\mathbf{V}_{k}}P(\mathbf{V}_k)\\
&=\int_{-\infty}^{\infty}\hspace{-1em}\cdots \int_{-\infty}^{\infty}\hspace{-1em}d\mathbf{V}_k\cdots d\mathbf{V}_1 \,\rho_{\mathbf{V}_{k}}P(\mathbf{V}_k,\dots,\mathbf{V}_1),\nonumber
\end{align}
where we have applied the definition of a marginal probability distribution, being $P(\mathbf{V}_k,\dots,\mathbf{V}_1)$ the joint distribution. Then, this joint distribution can be decomposed in terms of the conditional probability of the last event, such that $P(\mathbf{V}_k,\dots,\mathbf{V}_1)=P(\mathbf{V}_k|\mathbf{V}_{k-1},\dots,\mathbf{V}_1)P(\mathbf{V}_{k-1},\dots,\mathbf{V}_1)$. We are interested in this factorization because of the denominator of $\rho_{\mathbf{V}_{k}}$ in Eq. \eqref{eq:rhotrajectories}. 
The term $\textnormal{Tr}\left(\left(\mathcal{M}_{\mathbf{V}_k} \circ  \mathcal{L}_k\right)[\rho_{\mathbf{V}_{k-1}} ]\right)$ represents the probability of obtaining the outcome $\mathbf{V}_k$ given the previous history of $\rho_{\mathbf{V}_{k-1}}$. If we wrote this expression only in terms of the outcomes $\mathbf{V}_i$, this would be in fact the conditional probability $P(\mathbf{V}_k|\mathbf{V}_{k-1},\dots,\mathbf{V}_1)$. Gathering these definitions, we can finally derive Eq.~\eqref{eq:rhok}:
\begin{align} \label{eq:E2}
&\rho_k =\int_{-\infty}^{\infty}\hspace{-1em}\cdots \int_{-\infty}^{\infty}\hspace{-1em}d\mathbf{V}_k\cdots d\mathbf{V}_1 \left(\mathcal{M}_{\mathbf{V}_k} \circ  \mathcal{L}_k\right)[\rho_{\mathbf{V}_{k-1}}]\nonumber \\
&\frac{P(\mathbf{V}_k|\mathbf{V}_{k-1},\dots,\mathbf{V}_1)P(\mathbf{V}_{k-1},\dots,\mathbf{V}_1)}{P(\mathbf{V}_k|\mathbf{V}_{k-1},\dots,\mathbf{V}_1)} \nonumber \\
&=\int_{-\infty}^{\infty}\hspace{-1em}d\mathbf{V}_k\left(\mathcal{M}_{\mathbf{V}_k} \circ  \mathcal{L}_k\right) \nonumber \\
&\left[\int_{-\infty}^{\infty}\hspace{-1em}\cdots \int_{-\infty}^{\infty}\hspace{-1em}d\mathbf{V}_{k-1}\cdots d\mathbf{V}_1\, \rho_{\mathbf{V}_{k-1}}P(\mathbf{V}_{k-1},\dots,\mathbf{V}_1)\right] \nonumber \\
&=\int_{-\infty}^{\infty}\hspace{-1em}d\mathbf{V}_k\,\left(\mathcal{M}_{\mathbf{V}_k} \circ  \mathcal{L}_k\right)[\rho_{k-1}].
\end{align}
In the second to last equality, we have used the linearity of the map $\left(\mathcal{M}_{\mathbf{V}_k} \circ  \mathcal{L}_k\right)$ to introduce the integrals up to $\mathbf{V}_{k-1}$ into the argument, and then write $\rho_{k-1}$ with the definition of Eq.~\eqref{eq:E1}.

Next, we derive Eqs.~\eqref{eqrhoidealwithg} and \eqref{eq:M}. We begin from the last line of Eq.~\eqref{eq:E2}, introducing the measurement operators of Eq.~\eqref{eq:ourMk}:
\begin{equation}\label{eq:E3}
\rho_k=\int_{-\infty}^{\infty}\hspace{-1em}d\mathbf{V}_k \hat{\Phi}_{\mathbf{V}_k} \mathcal{L}_k[\rho_{k-1}] \hat{\Phi}^{\dagger}_{\mathbf{V}_k}=\int_{-\infty}^{\infty}\hspace{-1em}d\mathbf{V}_k \hat{\Phi}_{\mathbf{V}_k} \tilde{\rho}_{k} \hat{\Phi}^{\dagger}_{\mathbf{V}_k},
\end{equation}
where we just simplified notation with $\mathcal{L}_k[\rho_{k-1}]\equiv\tilde{\rho}_{k}$. Equation~\eqref{eq:E3} is a compact way of writing the integral of a matrix, but what we are really computing is the following: 
\begin{equation}\label{eq:E4}
\left[\rho_k\right]_{ij}=\int_{-\infty}^{\infty}\hspace{-1em}d\mathbf{V}_k \left[\hat{\Phi}_{\mathbf{V}_k} \tilde{\rho}_{k} \hat{\Phi}^{\dagger}_{\mathbf{V}_k}\right]_{ij},
\end{equation}
where indices $ij$ represent the matrix indices. Now, using Eq.~\eqref{eq:operV} for all qubits, i.e. in the z direction, it can be shown that the Kraus operator $\hat{\Phi}_{\mathbf{V}_k}$ can be written as
\begin{equation}\label{eq:E5}
    \hat{\Phi}_{\mathbf{V}_k}=\sum_if_i(\mathbf{V}_k)\ket{i}\bra{i},
\end{equation}
which is a diagonal operator in the z basis. Here $f_i(\mathbf{V}_k)$ is a product of $N$ gaussian distributions (see Eq.~\eqref{eq:operV}), where $N$ is the number of qubits and each gaussian is associated to the outcome of each spin. For example, for a system of $N=2$ qubits, the function associated to the basis state $\ket{00}$ is $f=(2\pi)^{-1/2}e^{-\frac{(V^{(0)}_k-g)^2}{4}}e^{-\frac{(V^{(1)}_k-g)^2}{4}}$.
By applying the measurement operators to the density matrix, we obtain that the matrix elements of the integral are 
\begin{equation}
  \left[\hat{\Phi}_{\mathbf{V}_k} \tilde{\rho}_{k} \hat{\Phi}^{\dagger}_{\mathbf{V}_k}\right]_{ij}=f_i(\mathbf{V}_k)f_j(\mathbf{V}_k) \left[\tilde{\rho}_{k}\right]_{ij}.
\end{equation}
Coming back to Eq.~\eqref{eq:E4},
\begin{equation}
\left[\rho_k\right]_{ij}=\left(\int_{-\infty}^{\infty}\hspace{-1em}d\mathbf{V}_k f_i(\mathbf{V}_k)f_j(\mathbf{V}_k)\right) \left[\tilde{\rho}_{k}\right]_{ij}=M_{ij}\left[\tilde{\rho}_{k}\right]_{ij}.
\end{equation}
We define the matrix $M$ such that its matrix elements are $M_{ij}\equiv\int_{-\infty}^{\infty} d\mathbf{V}_k f_i(\mathbf{V}_k)f_j(\mathbf{V}_k)$. By using matrix notation, we recover Eq.~\eqref{eqrhoidealwithg}:
\begin{equation}
    \rho_k=M\odot \tilde{\rho}_{k}=M\odot \mathcal{L}_k[\rho_{k-1}].
\end{equation}
 As we mentioned before, the functions $f_i(\mathbf{V}_k)$ are factorized in gaussians that are associated to each spin, so they can be written as $f_i(\mathbf{V}_k)=\prod^N_l \bar{f}_{il}(V^{(l)}_k)$, where $\bar{f}_{il}$ represents the contribution of  each qubit for the basis element $\ket{i}$. Then, matrix elements $M_{ij}$ can be obtained as:
\begin{equation}\label{eq:Mij}
    M_{ij}= \prod^N_l \int_{-\infty}^{\infty}\hspace{-1em}dV^{(l)}_k \bar{f}_{il}(V^{(l)}_k) \bar{f}_{jl}(V^{(l)}_k). 
\end{equation}
Each $\bar{f}_{il}(V^{(l)}_k)$ can only have two possible values, either $(2\pi)^{-1/4}e^{-\frac{(V^{(l)}_k-g)^2}{4}}$ or $(2\pi)^{-1/4}e^{-\frac{(V^{(l)}_k+g)^2}{4}}$. Then, there are only two possible outcomes for each integral: \begin{equation}
    \int_{-\infty}^{\infty}\hspace{-1em}dV^{(l)}_k \bar{f}_{il}(V^{(l)}_k) \bar{f}_{jl}(V^{(l)}_k)=\left\{\begin{matrix}
1 \quad \text{if } i=j \\
e^{-\frac{g^2}{2}} \quad \text{if } i\neq j.
\end{matrix}\right.
\end{equation}
The previous equation allows us to define the matrix $M$ of one single qubit, that we will denote as $\tilde{M}$. In such case, the measurement effect is represented by
\begin{equation}\label{eq:Mtilde}
    \tilde{M}=\begin{pmatrix}
1&e^{-\frac{g^2}{2}}\\
e^{-\frac{g^2}{2}}&1
\end{pmatrix}.
\end{equation}
Taking a look at Eq.~\eqref{eq:Mij}, we can observe that actually $M_{ij}$ is the result of a tensor product of single qubit matrices $\tilde{M}$, writing it as $M=\tilde{M}\otimes \cdots\otimes \tilde{M}=\tilde{M}^{\otimes N}$.

As a remark, we can apply these results for measuring in any direction, not only the z direction. We just need to rotate the density matrix towards the direction where the measurement is represented by the diagonal operator of Eq.~\eqref{eq:E5}. For example, measuring in the x direction, we apply the Hadamard gates before (for the measurement) and after (to return to the z basis):
\begin{equation}
    \rho_k = H \left(M\odot \left(H\mathcal{L}_k[\rho_{k-1}]H\right)\right) H
\end{equation}

Let us finally derive Eq.~\eqref{eqobsvsskandg}. Our goal is to see which $g$-factors could appear in the expression of the observables for a given input at time step $k$. We start from the definition of expected value:
\begin{equation}
    \langle \hat{\mathcal{O}} \rangle_{\rho_k}^{\infty}=\text{Tr}\left(\hat{\mathcal{O}}\rho_k\right)=\text{Tr}\left(\hat{\mathcal{O}}(M\odot \tilde{\rho}_{k})\right)=\text{Tr}\left(\hat{\mathcal{O}} \tilde{\rho}_{k}\right).
\end{equation}
The last equality is only true when we compute the expected value of an observable $\hat{\mathcal{O}}$ in the same direction where the measurements are performed. In this situation, $\hat{\mathcal{O}}$ is diagonal so the only matrix elements that contribute to the trace are the diagonal elements of $\tilde{\rho}_{k}$. As one can infer from Eq.~\eqref{eq:Mtilde}, the diagonal of the matrix $M$ is full of ones, so there is no back-action contribution at this level. This can be explained such that the measurements at time step $k$ do not introduce information about the back-action at that moment. Once the quantum state is updated with the measurement back-action, the effect will be observed at latter time steps. For this reason, we need to inspect at least one previous time step in the past to find an effect of the back-action (see Appendix~\ref{MethodsSec} for details on the reservoir map $\mathcal{L}_k$):
\begin{equation}
     \langle \hat{\mathcal{O}} \rangle_{\rho_k}^{\infty}=\text{Tr}\left(\hat{\mathcal{O}} \hat{U} \left(\rho_k^{\textnormal{in}}\otimes\textnormal{Tr}_{\textnormal{in}}\left[M\odot\tilde{\rho}_{k-1}\right]\right) \hat{U}^{\dagger}\right).
\end{equation}
For computing the partial trace, we need to write the matrices in blocks. We will use $2\times 2$ matrices composed by blocks of dimension $2^{N-1}\times 2^{N-1}$:
\begin{equation}
    \begin{split}
        &M=\tilde{M}^{\otimes N}=\begin{pmatrix}
\tilde{M}^{\otimes N-1}&e^{-\frac{g^2}{2}}\tilde{M}^{\otimes N-1}\\
e^{-\frac{g^2}{2}}\tilde{M}^{\otimes N-1}&\tilde{M}^{\otimes N-1}
\end{pmatrix}\\
&\tilde{\rho}_{k-1}=\begin{pmatrix}
\tilde{\rho}_{k-1}^{(1)}&\tilde{\rho}_{k-1}^{(2)}\\
\tilde{\rho}_{k-1}^{(3)}&\tilde{\rho}_{k-1}^{(4)}
\end{pmatrix}.
    \end{split}
\end{equation}
Then, the element-wise product gives us
\begin{equation}
M\odot \tilde{\rho}_{k-1}=\begin{pmatrix}
\tilde{M}^{\otimes N-1}\odot \tilde{\rho}_{k-1}^{(1)} &e^{-\frac{g^2}{2}}\tilde{M}^{\otimes N-1}\odot \tilde{\rho}_{k-1}^{(2)}\\
e^{-\frac{g^2}{2}}\tilde{M}^{\otimes N-1}\odot \tilde{\rho}_{k-1}^{(3)}&\tilde{M}^{\otimes N-1}\odot \tilde{\rho}_{k-1}^{(4)}
\end{pmatrix}.
\end{equation}
The partial trace is performed over the first qubit, so it can be computed as the sum of the diagonal blocks:
\begin{equation}
\textnormal{Tr}_{\textnormal{in}}\left[M\odot\tilde{\rho}_{k-1}\right]=\tilde{M}^{\otimes N-1}\odot ( \tilde{\rho}_{k-1}^{(1)}+\tilde{\rho}_{k-1}^{(4)}).
\end{equation}
As a consequence, every time step the measurement is applied over all the qubits, the factor $e^{-\frac{Ng^2}{2}}$ does not appear in the next time step because of the partial trace operation. Finally, applying an analysis similar to \cite{Mujal2021nonlinearities}, i.e. decomposing the input injection in three terms like $\rho_k^{\textnormal{in}}\otimes\textnormal{Tr}_{\textnormal{in}}\left[M\odot\tilde{\rho}_{k-1}\right]=A+Bs_{k}+Cr_k$, where  $r_k\equiv\sqrt{s_k(1-s_k)}$, we can obtain Eq.~\eqref{eqobsvsskandg}:
\begin{equation}
    \langle \hat{\mathcal{O}} \rangle_{\rho_k}^{\infty}=\sum_{n=1}^{N}e^{-\frac{N-n}{2}g^2}\left(A_{n}+B_{n}s_{k}+C_{n}r_k\right),
\end{equation}
with terms $A_n$, $B_n$ and $C_n$ gathering all the information about both the unitary dynamics $\hat{U}$ and the quantum state $\tilde{\rho}_{k-1}$ (which contains the information about the input sequence). All the combinations of exponential factors (except the term $e^{-\frac{Ng^2}{2}}$ due to the partial trace) appear as a consequence of the mixing of matrix elements of $\tilde{\rho}_{k-1}$ generated by the unitary dynamics.
\section{Measurements formalism}
\label{Appendix_Measurements_Formalism}
\subsection{Single-qubit measurements}
\label{appendix_single_qubit_measurements}
\subsubsection{Measurements in the x and y directions}
In addition to the measurement operator in the z direction given in the main text, in Eq.~\eqref{eq:operV}, here we write the analogous expressions for the x and y directions, respectively:
\begin{align}
    &\hat{\Omega}^{\rm x}_V \equiv \hat{H}_{\rm D} \hat{\Omega}_V \hat{H}_{\rm D}=
\\
&\frac{1}{(32\pi)^{1/4}}\left[\left(e^{-\frac{(V-g)^2}{4}}+ e^{-\frac{(V+g)^2}{4}}\right)\left(\ket{0}\bra{0}+\ket{1}\bra{1}\right)\right. \nonumber
\\
&\left. +\left(e^{-\frac{(V-g)^2}{4}}- e^{-\frac{(V+g)^2}{4}}\right)\left(\ket{0}\bra{1}+\ket{1}\bra{0}\right)\right],\nonumber
\end{align}
and,
\begin{align}
&\hat{\Omega}^{\rm y}_V \equiv \left(\hat{H}_{\rm D}\hat{S}^{\dagger}\right)^{\dagger} \hat{\Omega}_V \hat{H}_{\rm D}\hat{S}^{\dagger}=
\\
&\frac{1}{(32\pi)^{1/4}}\left[\left(e^{-\frac{(V-g)^2}{4}}+ e^{-\frac{(V+g)^2}{4}}\right)\left(\ket{0}\bra{0}+\ket{1}\bra{1}\right)\right.\nonumber
\\
&\left. -i \left(e^{-\frac{(V-g)^2}{4}}- e^{-\frac{(V+g)^2}{4}}\right)\left(\ket{0}\bra{1}-\ket{1}\bra{0}\right)\right],\nonumber
\end{align}
where $\hat{H}_{\rm D}$ is the Hadamard operator, $\hat{H}_{\rm D}=(1/\sqrt{2})\left(\ket{0}\bra{0}+\ket{0}\bra{1}+\ket{1}\bra{0}-\ket{1}\bra{1}\right)$, and $\hat{S}$ is the phase shift operator, $\hat{S}=\ket{0}\bra{0}+i\ket{1}\bra{1}$.

The corresponding probability distributions (analogous to Eq.~\eqref{eq:P_V}) of an outcome $V$ for a reduced single-qubit state $\omega$ to the previous two equations are, respectively, given by:
\begin{align}
    &P_{\rm x}(V)=\text{Tr}((\hat{\Omega}^{\rm x}_V)^{\dagger}\hat{\Omega}^{\rm x}_V\omega)=
    \\
    &\frac{1}{\sqrt{2\pi}}\left[\left(\frac{1}{2}+\Re(\omega_{01})\right)e^{-\frac{(V-g)^2}{2}}+\left(\frac{1}{2}-\Re(\omega_{01})\right)e^{-\frac{(V+g)^2}{2}}\right],\nonumber
    \end{align}
and,
\begin{align}
    &P_{\rm y}(V)=\text{Tr}((\hat{\Omega}^{\rm y}_V)^{\dagger}\hat{\Omega}^{\rm y}_V\omega)=
    \\
    &\frac{1}{\sqrt{2\pi}}\left[\left(\frac{1}{2}-\Im(\omega_{01})\right)e^{-\frac{(V-g)^2}{2}}+\left(\frac{1}{2}+\Im(\omega_{01})\right)e^{-\frac{(V+g)^2}{2}}\right],\nonumber
\end{align}
where $\Re()$ and $\Im()$, indicate, respectively, the real and imaginary part of a complex number.
\subsubsection{Expectation values and uncertainties}
From the last equations, we also derive the analogous expressions to $   \langle V \rangle_{\omega}^{\rm z}=\int_{-\infty}^{\infty} V \, P_{\rm z}(V)\, dV=g\langle \hat{\sigma}^{\rm z} \rangle_{\omega},$ for the x and y directions. The expectation value of $V$ when the measurements are performed in the x and y directions, are, respectively, for the state $\omega$, given by:
\begin{equation}
    \langle V \rangle^{\rm x}_{\omega}=\int_{-\infty}^{\infty}\hspace{-1em}V \, P_{\rm x}(V)\, dV=g\langle \hat{\sigma}^{\rm x} \rangle_{\omega},
\end{equation}
and
\begin{equation}
    \langle V \rangle^{\rm y}_{\omega}=\int_{-\infty}^{\infty}\hspace{-1em}V \, P_{\rm y}(V)\, dV=g\langle \hat{\sigma}^{\rm y} \rangle_{\omega},
\end{equation}
with $\langle \hat{\sigma}^{\rm x} \rangle_{\omega}=\text{Tr}(\omega \hat{\sigma}^{\rm x})$, and $\langle \hat{\sigma}^{\rm y} \rangle_{\omega}=\text{Tr}(\omega \hat{\sigma}^{\rm y})$.

Getting back to the case of measurements in the z direction, we compute the uncertainty associated to the expectation value by additionally computing
\begin{equation}
    \langle V^2 \rangle_{\omega}^{\rm z}=\int_{-\infty}^{\infty}\hspace{-1em}V^2 \, P_{\rm z}(V)\, dV=g^2+1.
\end{equation}
Then, the dispersion reads:
\begin{equation}
s_V^{\rm z}=\sqrt{\langle V^2 \rangle_{\omega}^{\rm z}-(\langle V \rangle_{\omega}^{\rm z})^2}=\sqrt{1+g^2 s_{z,\omega}^2},
\end{equation}
with, $s_{z,\omega}^2\equiv 1-\langle \hat{\sigma}^{\rm z} \rangle_{\omega}^2=1-(2\omega_{00}-1)^2$. The standard deviation of the mean value estimated with $N_{\rm meas}$ measurement results, namely, as $\langle V \rangle_{\omega}^{\rm z}\approx(1/N_{\rm meas})\sum_{l=1}^{N_{\rm meas}}V_l^{\rm z}$, is
\begin{equation}
\bar{s}_V^{\rm z}=\frac{s_V^{\rm z}}{\sqrt{N_{\rm meas}}}=\sqrt{\frac{1+g^2 s_{z,\omega}^2}{N_{\rm meas}}}.
\end{equation}
Finally, the state-dependent uncertainty is computed as
\begin{equation}
\label{eq_uncertainty_statedependent}
    \bar{s}_{\langle\hat{\sigma}^{\rm z}\rangle_{\omega}}=\frac{\bar{s}_V^{\rm z}}{g}=\sqrt{\frac{1+g^2 s^2_{z,\omega}}{g^2N_{\rm meas}}}.
\end{equation}
For the x and y directions, the same analogous expressions are found for $\bar{s}_{\langle\hat{\sigma}^{\rm x}\rangle_{\omega}}$ and $\bar{s}_{\langle\hat{\sigma}^{\rm y}\rangle_{\omega}}$ by replacing $s^2_{z,\omega}$ with $s^2_{x,\omega}$ and $s^2_{y,\omega}$, respectively. The state-independent and direction-independent uncertainty given in Eq.~\eqref{eqstatisticalsinglequbits} is the upper bound of $\bar{s}_{\langle\hat{\sigma}^{\rm x}\rangle_{\omega}}$, $\bar{s}_{\langle\hat{\sigma}^{\rm y}\rangle_{\omega}}$, and $\bar{s}_{\langle\hat{\sigma}^{\rm z}\rangle_{\omega}}$. Namely,
\begin{equation}
\label{equpperbounduncertainty}
    \bar{s}_{\langle\hat{\sigma}^{\alpha}\rangle_{\omega}} \leq \bar{s}_{\langle\hat{\sigma}\rangle}=\sqrt{\frac{1+g^2}{g^2N_{\text{meas}}}},
\end{equation}
which is valid for any reduced single-qubit state $\omega$ and direction $\alpha=$ x, y, z. This maximum uncertainty is reached in the direction $\alpha$ when $\langle \hat{\sigma}^{\alpha} \rangle_{\omega}=0$.

The previous equation is useful to compare a measurement with strength $g$ to a measurement with strength $g'$. In order to estimate an expectation value with the same statistical uncertainty, namely, $\bar{s}_{\langle\hat{\sigma}\rangle}=\bar{s}'_{\langle\hat{\sigma}\rangle}$, the number of measurements associated to each case, $N_{\rm meas}$ and $N_{\rm meas}'$, respectively, are related as follows:
\begin{equation}
\label{eqNmesNmeas}
N_{\rm meas}=N_{\rm meas}'\frac{(1+g^2)g'^2}{(1+g'^2)g^2}.
\end{equation}
For $g<g'$, in general, more weaker measurements are needed, $N_{\rm meas}>N_{\rm meas}'$. In particular, we can compare projective measurements with weaker ones. In the limit of $g \ll g'$ we have:
\begin{equation}
\label{eqNmesNmeaslargeg}
N_{\rm meas}=N_{\rm meas}'\left(1+\frac{1}{g^2}\right).
\end{equation}

The upper bound in Eq.~\eqref{equpperbounduncertainty} is closely reached in some limits. We can write the expression in \eqref{eq_uncertainty_statedependent} for the direction $\alpha$ as
\begin{equation}
   \bar{s}_{{\langle\hat{\sigma}^{\alpha}\rangle}_{\omega}}=\sqrt{\frac{1+g^2}{g^2N_{\text{meas}}}}\sqrt{1-\epsilon}, 
\end{equation}
where $\epsilon={\langle\hat{\sigma}^{\alpha}\rangle}^2_{\omega}/(1+1/g^2)$. In the limits of either $g\rightarrow 0$ or ${\langle\hat{\sigma}^{\alpha}\rangle_{\omega}}\rightarrow 0$, we will have that $\epsilon \rightarrow 0$. Then, we can approximate the previous equation as 
\begin{equation}
   \bar{s}_{{\langle\hat{\sigma}^{\alpha}\rangle}_{\omega}}\simeq \sqrt{\frac{1+g^2}{g^2N_{\text{meas}}}}(1-\frac{\epsilon}{2}). 
\end{equation}
In these two situations, the dependence on the state and on the direction are small first-order corrections to the upper bound, which is the main zeroth-order term.
\subsection{Two-qubit measurements}
\label{appendix_twoqubit_measurements}
\subsubsection{Expectation values and uncertainties}
In this section, we consider a reduced two-qubit state $\omega$, with components $\omega_{ij}$, with $i,j\in\{0,1,2,3\}$. Our goal is to measure the expectation value of two-qubit observables of the form $\hat{\sigma}^\alpha\otimes\hat{\sigma}^\alpha$ through indirect measurements. In the following lines, as an example, we treat the case of the operator $\hat{\sigma}^{\rm z}\otimes\hat{\sigma}^{\rm z}$.

The two-qubit indirect measurement operator in consideration is $\hat{\Omega}_{V_1}^{\rm z}\otimes \hat{\Omega}_{V_2}^{\rm z}$, which is a product of two single-qubit operators whose form is given in Eq.~(\ref{eq:operV}). The probability of obtaining the measurement outcomes $V_1$ and $V_2$ with a measurement on the state $\omega$ reads:
\begin{align}
&P_{\rm z}(V_1,V_2)=\textnormal{Tr}\left[(\hat{\Omega}_{V_1}^{\rm z}\otimes\hat{\Omega}_{V_2}^{\rm z})^\dagger(\hat{\Omega}_{V_1}^{\rm z}\otimes\hat{\Omega}_{V_2}^{\rm z})\omega\right]=
\\
&\frac{1}{2\pi}\left(\omega_{00}e^{-\frac{(V_1-g)^2}{2}}e^{-\frac{(V_2-g)^2}{2}}+\omega_{11}e^{-\frac{(V_1-g)^2}{2}}e^{-\frac{(V_2+g)^2}{2}}+\right.\nonumber
\\
&\left.\omega_{22}e^{-\frac{(V_1+g)^2}{2}}e^{-\frac{(V_2-g)^2}{2}}+\omega_{33}e^{-\frac{(V_1+g)^2}{2}}e^{-\frac{(V_2+g)^2}{2}}\right).\nonumber
\end{align}
From this last expression, we compute the mean value of the product of the measurement results:
\begin{equation}
\begin{split}
    &\langle V_1 V_2\rangle_{\omega}^{\rm z}=\int_{-\infty}^{\infty}\hspace{-1em}dV_1\,\int_{-\infty}^{\infty}\hspace{-1em}dV_2\,P_{\rm z}(V_1,V_2) V_1 V_2 =
    \\
    &g^2\langle\hat{\sigma}^{\rm z}\otimes\hat{\sigma}^{\rm z} \rangle_{\omega},
    \end{split}
\end{equation}
and also the statistical uncertainty:
\begin{equation}
\begin{split}
&s_{(V_1V_2)}^{\rm z}=\sqrt{\langle (V_1 V_2)^2 \rangle_{\omega}^{\rm z}-(\langle V_1 V_2 \rangle_{\omega}^{\rm z})^2}=
\\
&\sqrt{1+2g^2+g^4(1-\langle\hat{\sigma}^{\rm z}\otimes\hat{\sigma}^{\rm z} \rangle^2_{\omega})},
\end{split}
\end{equation}
where we have used that:
\begin{equation}
    \langle V_1^2 V_2^2\rangle^{\rm z}=\int_{-\infty}^{\infty}\hspace{-1em}dV_1\,\int_{-\infty}^{\infty}\hspace{-1em}dV_2\,P_{\rm z}(V_1,V_2) V_1^2 V_2^2 =(g^2+1)^2.
\end{equation}
As before for the single-qubit case, we similarly calculate the standard deviation of the mean value estimated with $N_{\rm meas}$ measurements, namely, with $\langle V_1 V_2 \rangle_{\omega}^{\rm z}\approx(1/N_{\rm meas})\sum_{l=1}^{N_{\rm meas}}(V_1 V_2)_{l}^{\rm z}$,
\begin{equation}
\bar{s}_{(V_1V_2)}^{\rm z}=\frac{s_{(V_1V_2)}^{\rm z}}{\sqrt{N_{\rm meas}}}=\sqrt{\frac{1+2g^2+g^4(1-\langle\hat{\sigma}^{\rm z}\otimes\hat{\sigma}^{\rm z} \rangle^2_{\omega})}{N_{\rm meas}}},
\end{equation}
and for $\langle\hat{\sigma}^{\rm z}\otimes\hat{\sigma}^{\rm z} \rangle_{\omega}$, the associated uncertainty is given by:
\begin{equation}
\bar{s}_{\langle\hat{\sigma}^{\rm z}\otimes\hat{\sigma}^{\rm z}\rangle_{\omega}}=\frac{\bar{s}_{(V_1V_2)}^{\rm z}}{g^2}=\sqrt{\frac{1+2g^2+g^4(1-\langle\hat{\sigma}^{\rm z}\otimes\hat{\sigma}^{\rm z} \rangle^2_{\omega})}{g^4 N_{\rm meas}}},
\end{equation}
As well as in Eq.~\eqref{equpperbounduncertainty}, in the main text, in Eq.~\eqref{eqstatisticaltwoqubits}, we consider the upper bound of the statistical uncertainty, which is independent of both the state and the direction of measurement,
\begin{equation}
\label{equpperbounduncertainty_twoqubits}
    \bar{s}_{\langle\hat{\sigma}^{\alpha}\otimes \hat{\sigma}^{\alpha}\rangle_{\omega}} \leq \bar{s}_{\langle\hat{\sigma}\otimes\hat{\sigma}\rangle}=\sqrt{\frac{1+2g^2+g^4}{g^4 N_{\rm meas}}},
\end{equation}
for any reduced two-qubit state $\omega$ and direction $\alpha=$ x, y, z.

As before, this last equation can be used to compare a measurement with strength $g$ to a measurement with strength $g'$. In order to estimate an expectation value with the same statistical uncertainty, namely, $\bar{s}_{\langle\hat{\sigma}\otimes\hat{\sigma}\rangle}=\bar{s}'_{\langle\hat{\sigma}\otimes\hat{\sigma}\rangle}$, the number of measurements associated to each case, $N_{\rm meas}$ and $N_{\rm meas}'$, respectively, are related as follows:
\begin{equation}
\label{eqNmesNmeastwoqubit}
N_{\rm meas}=N_{\rm meas}'\frac{(1+2g^2+g^4)g'^4}{(1+2g'^2+g'^4)g^4}.
\end{equation}
\section{Quantum reservoir based on a disordered transverse Ising model}
\label{MethodsSec}

Reservoir computing~\cite{Book_Nakajima_Fischer2021,tanaka2019recent,konkoli2017reservoir} is a supervised machine learning method based on the three-layer scheme: (i) input, (ii) reservoir, and (iii) output, depicted in Fig.~\ref{fig:protocols}(a). The main enabling features of reservoir computing are the ability to differentiate any pair of inputs, known as separability \cite{grigoryeva2018universal}, the fading memory in the past and the independence of the reservoir initialization (echo state or convergence property) \cite{grigoryeva2018echo}. This guarantees that after a transient time the reservoir output depends only on the recent input sequence. Recent proposals for quantum reservoir computing have been reviewed in~\cite{mujal2021opportunities}.

The  quantum reservoir computer considered in this work consists of a qubit network evolving in time under the the action of the disordered transverse-field Ising Hamiltonian~\cite{Nakajima2017,martinez2021dynamical}:
\begin{equation}
\hat{H}=\frac{1}{2}\sum_{i=0}^{N-1}h\hat{\sigma}_i^{\rm z}+\sum_{i<j}^{N-1}J_{ij}\hat{\sigma}_i^{\rm x}\hat{\sigma}_j^{\rm x}.
\label{hamtransverseising}
\end{equation}
In all the numerical simulations, we used $N=6$ qubits and randomly generated {a single set of} qubit-qubit couplings $J_{ij}$ from a uniform distribution in the interval $[-J_s/2,J_s/2]$, where $J_s$ is set as a reference unit. {We have checked that the deviations introduced by different coupling realizations is not relevant in our analysis, as can be seen in the example of Appendix~\ref{networkrealizations}.} The external magnetic field and the time interval, respectively, are fixed at $h=10J_s$ and $\Delta t=10/J_s$ so that the reservoir is in an appropriate dynamical regime where the system tends to thermalize~\cite{martinez2021dynamical}. The input is represented  by a sequence
 $\{s_0,s_1,\dots,s_k,\dots\}$  and is injected into the system by rewriting a node state $\rho_k^{\textnormal{in}}$ every time step $k$. In particular, we consider encoding  a real input $s_k\in [0,1]$ in one qubit (labeled by $0$) by setting it at each time into a pure state $\ket{\psi_k}=\sqrt{1-s_k}\ket{0}+\sqrt{s_k}\ket{1} $ \cite{Mujal2021nonlinearities}.
The resulting evolution after input injection is given by the completely positive trace-preserving map $  \mathcal{L}_k[\rho]=\hat{U} \left(\rho_k^{\textnormal{in}}\otimes\textnormal{Tr}_{\textnormal{in}}\left[\rho\right]\right) \hat{U}^{\dagger}$ \cite{Nakajima2017,martinez2021dynamical}. This fully defines the unperturbed state dynamics in the case of the RSP and the RWP, as given by Eqs.~\eqref{eq:rhokunperturbed} and \eqref{eq:rhotrajectories}.

\subsection{QRC training and test}
For task resolution, the QRC systems are trained in a supervised manner.
Given the target output $y_k$ for the input $s_k$ in the training examples, a linear weighted combination of the reservoir observables is constructed to minimize the prediction error with a least-squares method \cite{lukovsevivcius2009reservoir}. 
More precisely, the form of the predictions is
\begin{equation}
\label{predictionsmodeleq}
    \tilde{y}_k= \sum_{m=1}^{L}w_m \langle\hat{\mathcal{O}}\rangle_{k,(m)}+ w_{L+1},
\end{equation}
where $w_m$ are the free parameters (weights) to be optimized, including the bias term $w_{L+1}$. The total number of weights employed, $L$, depends on our particular choice for the observable set. For instance, when we consider the expectation values of $\hat{\sigma}^{\rm x}_i$, $\hat{\sigma}^{\rm y}_i$, and $\hat{\sigma}^{\rm z}_i$ for all qubits, then $L=3N$.

For the STM task, we consider a total dataset of $N_t=1000$ time steps. 
The initial $N_{\rm wo}=20$ instants within the washout time are discarded. This value, in agreement with previous results~\cite{martinez2020information}, is long enough to guarantee a nearly identical performance between the RSP and the RWP.
It is also valid for the OLP even if a shorter time could be enough, as the memory of the system is reduced for increasing values of $g$ due to back-action.
The weights in Eq.~\eqref{predictionsmodeleq} are optimized using the following 735 time instants and the last 245 time steps constitute the test set, where the learned weights are used to evaluate the accuracy of the QRC system.

For the Santa Fe task, the data set is comprised of $N_t=2000$ time steps. The first $N_{\rm wo}=20$ time steps are again discarded. The following 70$\%$ instants are used to train the weights, while the test prediction error is evaluated for the last 30$\%$ steps.

The QRC perfomance is quantitatively evaluated through the capacity 
\begin{equation}
\label{eq:capacitydefinition}
    C \equiv \frac{{\rm cov}^2\left({\bm y},\tilde{\bm y}\right)}{{\rm var}({\bm y}){\rm var}(\tilde{{\bm y}})},
\end{equation}
where ${\rm cov}\left({\bm y},\tilde{\bm y}\right)$ indicates the covariance between the two series and ${\rm var}({\bm y})$ is the variance. $C\in[0,1]$ and perfect predictions are obtained with $C=1$.
\section{QRC performance considering different random qubit networks}
\label{networkrealizations}
\begin{figure}[t!]
    \centering
    \includegraphics[width=\columnwidth]{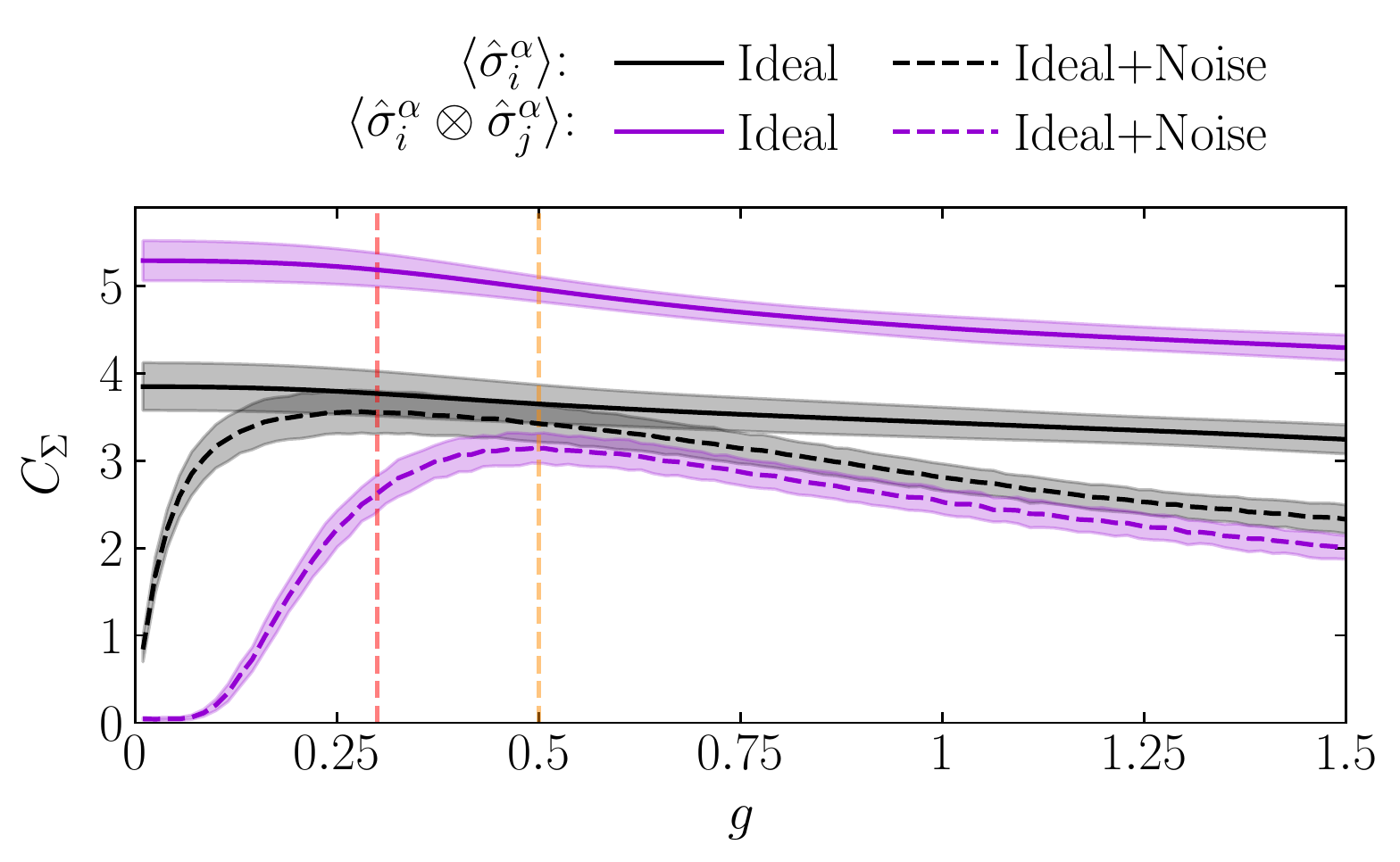}
    \caption{STM sum capacity depending on $g$ for different kind of observables in the OLP averaged over 100 different qubit-qubit couplings $J_{ij}$ realizations. Solid lines represent the ideal values when $N_{\rm meas}\rightarrow \infty$. Dashed lines corresponds to estimated values obtained again in the limit $N_{\rm meas}\rightarrow \infty$ but adding a Gaussian noise to the observables. The shaded regions correspond to the standard deviation over couplings' realizations. Vertical dashed lines represent the values of $g=0.3$ (red) and $g=0.5$ (orange). In the plot legend, $\alpha=$ x, y, z, and $i,j=1,\, ..., N=6$. All data shown correspond to the test set.}
    \label{fig:averages}
\end{figure}
All the results in the main text can be consistently reproduced when considering different realizations of the disordered Ising model in Eq.~\eqref{hamtransverseising}. In Fig.~\ref{fig:averages}, we show as an example the STM sum capacity depending on $g$ for single and two-qubit observables in the OLP, resembling Fig.~\ref{fig:sumcapacityvsnmeasandtexp}(a) of the main text. All the lines correspond to the ideal case of $N_{\rm meas}\rightarrow \infty$ with back-action, adding Gaussian noise (with standard deviation given by Eqs.~\eqref{eqstatisticalsinglequbits} and \eqref{eqstatisticaltwoqubits}) in the observables for the dashed lines. The shadowed region displays the standard deviation when considering a sample of 100 networks, with different qubit-qubit couplings $J_{ij}$. We see that no large variations in the capacity are observed for different random networks, confirming the same qualitative behaviour that was obtained in Fig.~\ref{fig:sumcapacityvsnmeasandtexp}(a) of the main text for a single realization.

We also note that while we have obtained the results in the main paper mostly considering monitored quantum trajectories, here instead we have estimated the effect of statistical noise due to the limited ensemble size by introducing a single realization of Gaussian noise. This stochastic fluctuation of zero mean and standard deviation given by Eqs.~\eqref{eqstatisticalsinglequbits} or \eqref{eqstatisticaltwoqubits} (depending on which observable we use) accounts for the largest deviation we could find in an experiment with a given measurement strength $g$ and number of measurements $N_{\rm meas}$.
\begin{figure}[t!]
    \centering
    \includegraphics[width=\columnwidth]{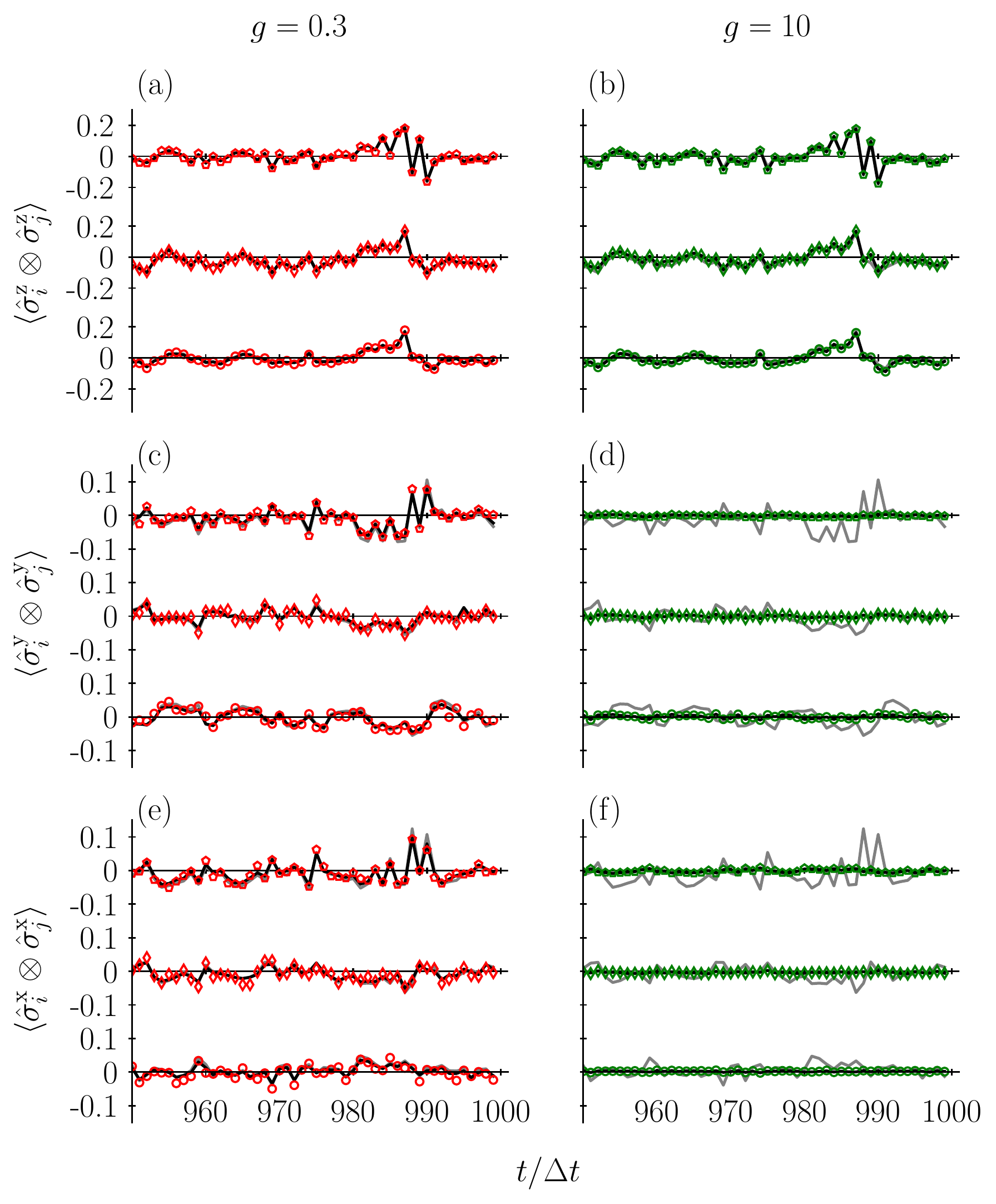}
    \caption{Two-qubit observables in the x, y and z directions. Solid lines correspond to the ideal values with the effect of measurements on the system (black) and the unperturbed case (grey), respectively. In symbols, the estimated values obtained by numerically simulating the measurement process with $N_{\rm meas}=1.5 \cdot 10^6$ by following the OLP are represented. Panels (a), (c), and (e) correspond to a weak measurement strength, $g=0.3$; whereas in panels (b), (d), and (f) strong measurements, $g=10$, are performed. The input is encoded into the state of qubit 0 (see Appendix~\ref{MethodsSec}). In each panel, from bottom to top, we plot the observables corresponding to the following pairs of qubits: $i=0$, $j=1$; $i=0$, $j=2$; and $i=1$, $j=2$.}
    \label{fig:twoqubitobservables}
\end{figure}
\section{QRC performance using second order moments}
\label{appendix_observables}
The output layer of QRC can be based on different observables and in the main body of the paper we illustrate the performance mostly considering first-order moments. Here, we provide more detailed results when considering two-qubit observables in the x, y and z directions. 
We start by showing the impact of weak/strong measurement on the two Pauli operators strings complementing  Fig.~\ref{fig:singlequbitobservables} in the main text.
In Fig.~\ref{fig:twoqubitobservables}, we see that also these observables follow the unperturbed dynamics for weak measurement, while qubit correlations in the x and y directions are erased due to strong measurement back-action, so that only the z components follow the target even when strongly monitored. We can explain again this behaviour because of a dynamics mostly aligned in the z-direction of the reservoir.
 
\begin{figure}[t!]
    \centering
    \includegraphics[width=\columnwidth]{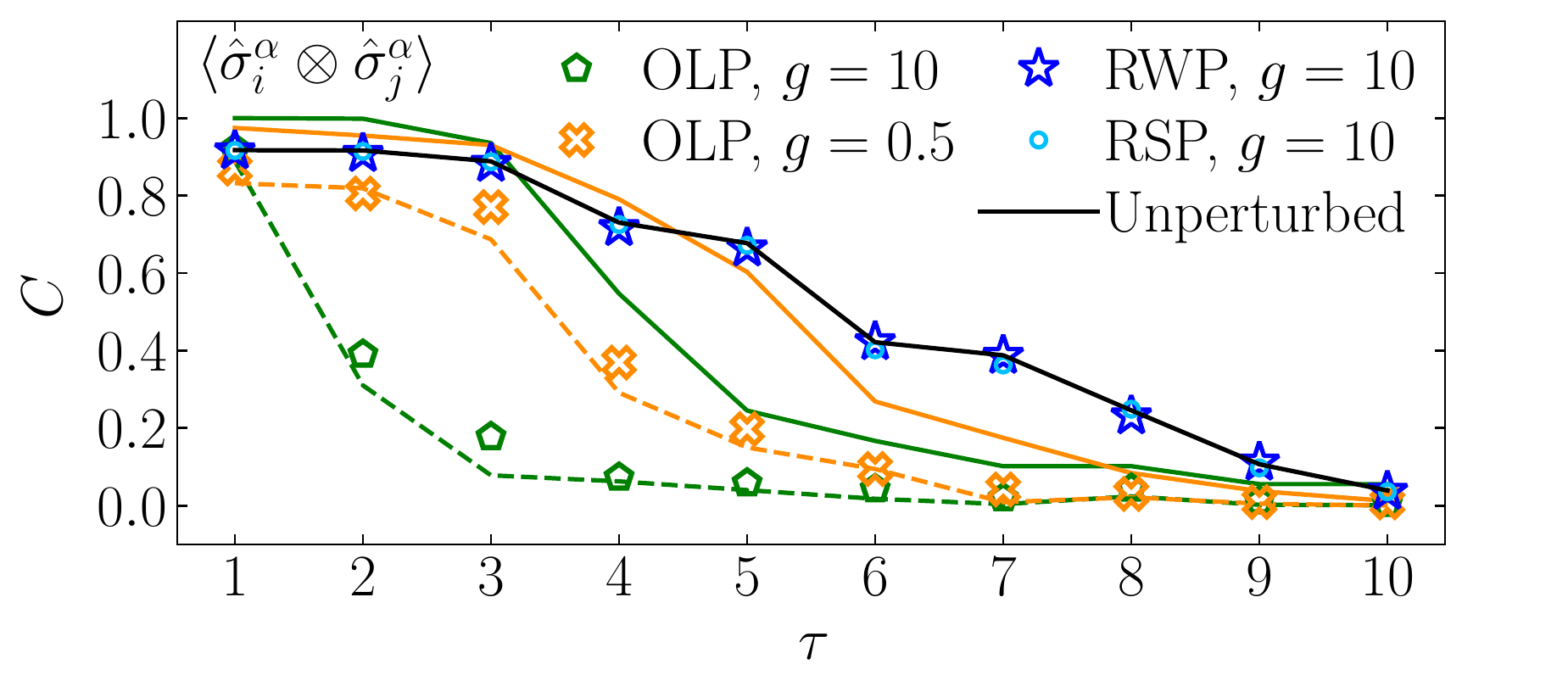}
    \caption{STM capacity from two-qubit observables depending on the delay $\tau$ obtained with the different protocols and $N_{\rm meas}=1.5\cdot 10^6$ measurements (symbols). The limit cases with $N_{\rm meas}\rightarrow \infty$, are represented with a solid line in the corresponding color; the unperturbed situation is in black. The dashed lines corresponds to estimated values obtained again in the limit $N_{\rm meas}\rightarrow \infty$ but adding a Gaussian noise to the observables. $i=0,\, ...,N=6$ and $\alpha=$ x, y, z. }
    \label{captautwoqubit}
\end{figure}

Figure~\ref{captautwoqubit} shows the STM capacity of the two-qubit observables against the delay of the target, similar to Fig.~\ref{fig:targetpred_capacities_vs_tau}(a) for single-qubit observables. However, in this case we study the OLP for $g=0.5$ instead of $g=0.3$. Here, the unperturbed case shows a memory tail that decays slower than for single-qubit observables, which is reasonable since we are using more observables. And again the RWP and RSP (blue symbols), for a finite number $N_{\rm meas}=1.5\cdot 10^6$ of measurements, reach the performance of the unperturbed case. With respect to the OLP, the measurement strength is critical, but the situation is different to the case of single-qubit observables. For $g=10$ (green symbols), there is an exponential memory decay in terms of $\tau$, while $g=0.5$ (orange symbols) preserves the memory plateau of the unperturbed case for small delays. But while we observed in Fig.~\ref{fig:targetpred_capacities_vs_tau}(a) that weak measurements preserved the whole tail of memory, here we do not see this behaviour for the longer delays. 

In the main text, we explained the hindered performance of $g=10$ as the result of both back-action and statistical noise, but the latter has now an important role for $g=0.5$. In fact, as can be seen in Eq.~\eqref{eqstatisticaltwoqubits},  now the dependence on $g$ makes necessary a much larger number of measurements to reduce the statistical error. Assuming infinite measurements (ideal cases with back-action, represented by colored continuous lines), we can isolate the effect of the back-action with respect to the statistics. In this situation, the performance is considerably improved, but still, the strong back-action effect of $g=10$ affects the memory of the reservoir for the longer delays. Applying a Gaussian noise with a standard deviation given by Eq.~\eqref{eqstatisticaltwoqubits} to these ideal cases, we recover the OLP with finite samples.
%
\begin{figure}[t!]
    \centering
    \includegraphics[width=\columnwidth]{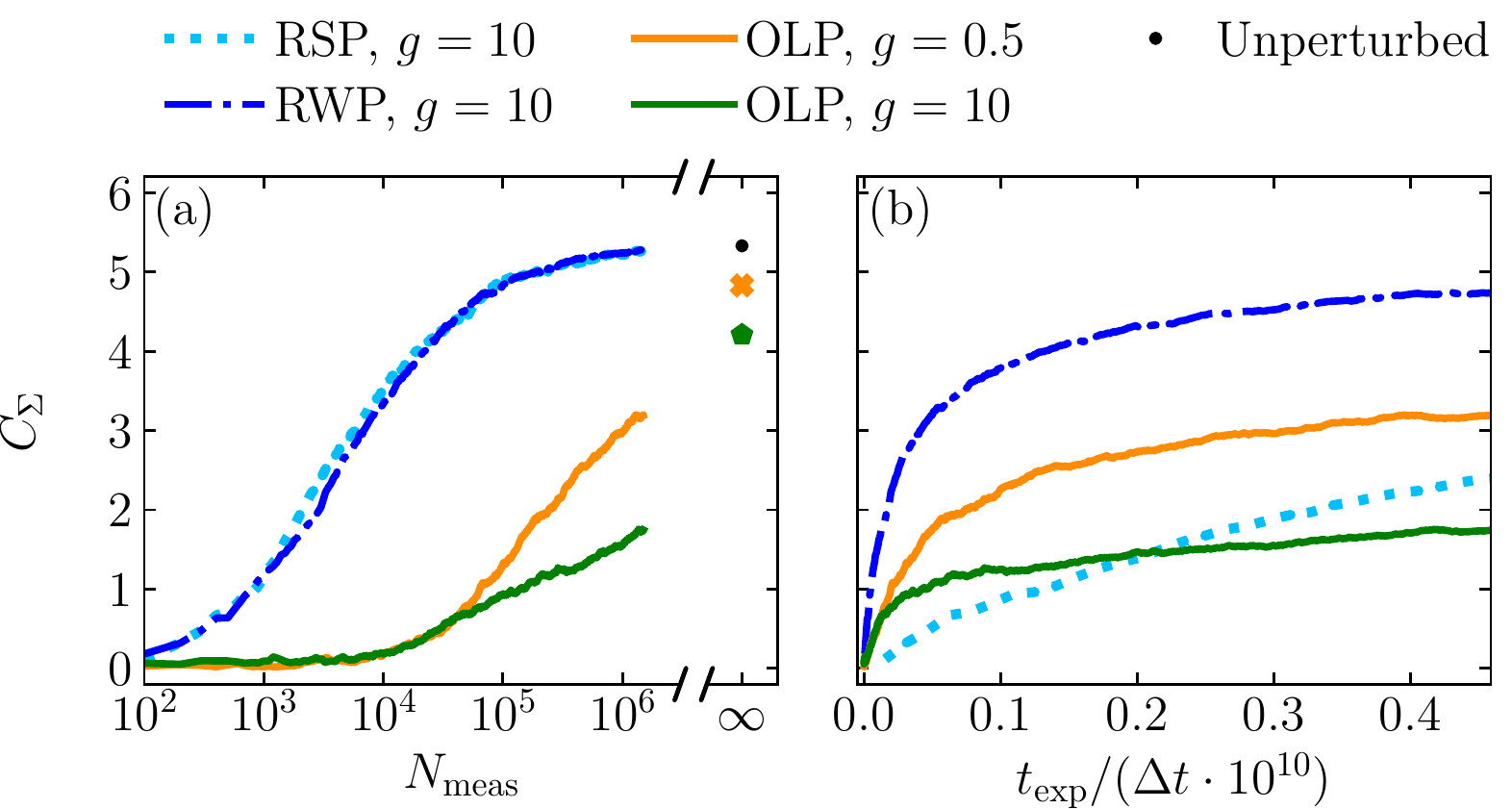}
    \caption{(a) STM sum capacity as a function of the number of measurements, and, (b) depending on the experimental time required. Two-qubit observables, $\langle \hat{\sigma}^{\alpha}_{i} \otimes \hat{\sigma}^{\alpha}_{j}\rangle$, are used with $\alpha=$ x, y, z and $i,j=1,\, ..., N=6$. All data shown correspond to the test set}
    \label{fig:sumcapacityvsnmeasandtexp_twoqubitobs}
\end{figure}

Finally, Fig.~\ref{fig:sumcapacityvsnmeasandtexp_twoqubitobs} is the two-qubit representation of Figs.~\ref{fig:sumcapacityvsnmeasandtexp}(b) and (c) of the main text. The STM capacity sum is represented as a function of the number of measurements for all the protocols, taking $g=0.5$ for the weak measurements of the OLP. In (a) we find that the sum capacities of the RSP and RWP increase almost to converge to the unperturbed case. But in this case, it is required to increase the number of measurements by almost one order of magnitude with respect to the single-qubit case to see this saturation. Accordingly, in the OLP the increase of the sum capacity is also displaced to larger values of $N_{\rm meas}$. As for single-qubit observables, $g=10$ in the OLP shows a slower convergence of the sum capacity. In contrast, $g=0.5$ shows a steeper slope, indicating the possibility of reaching the ideal sum capacity for a smaller $N_{\rm meas}$. Figure~\ref{fig:sumcapacityvsnmeasandtexp_twoqubitobs}(b) displays the STM sum capacity in terms of the experimental time to indicate the real needed resources by each protocol. As discussed in the main text, the RSP is very inefficient in comparison with the RWP. Now, the examples of OLP require a longer experimental time to reach the same sum capacity as in the RWP. This behavior indicates that the back-action of $g=0.5$ is strong enough to be detrimental for the performance.
\section{Memory capacity $C(\tau=1)$ and non-classical signatures of QRC} \label{appendix_nonclassical}

The maximum STM capacity is obtained when targeting the shortest possible delay, that is delay $\tau=1$, and it progressively decays with delay. In the main text, we have illustrated the ability of the system to reproduce the target for delay $\tau=2$ while here we add the plots in the more favorable case  $C(\tau=1)$. We also show separately the effect of measurement in all the directions.

\begin{figure}[t!]
    \centering
    \includegraphics[width=\columnwidth]{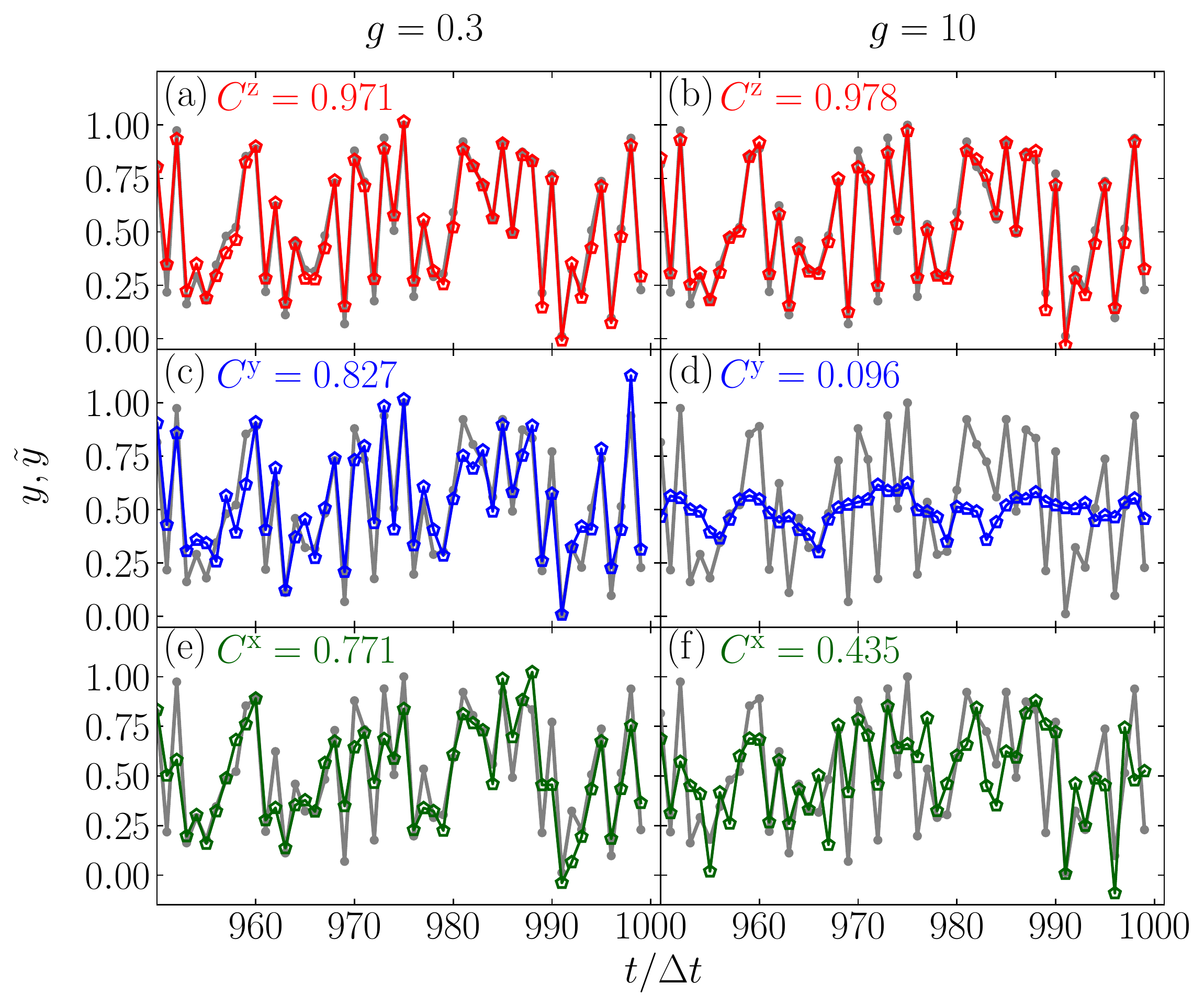}
    \caption{Target values (grey filled circles) and predictions (colored unfilled symbols) for the STM task at $\tau=1$ obtained with the OLP in different directions separately with weak measurements (left panels), and strong measurements (right panels), both with single-qubit observables. All data shown correspond to the test set.}
    \label{fig:detailsdirections}
\end{figure}

Figure~\ref{fig:detailsdirections} represents a time window of the target values (grey) and the predictions obtained for the projection observables (color) in directions x, y and z.  These predictions correspond to the OLP, where left panels are produced with weak measurements ($g=0.3$) and right panels are produced with strong measurements ($g=10$). For weak measurements (left panels (a), (c) and (e)), the contributions from all three directions are significant. However, in the strong measurement case (right panels (b), (d) and (f)), contributions from x and y directions are drastically reduced, in particular the y-direction. As in Figs.~\ref{fig:targetpred_capacities_vs_tau}(b) and (c) of the main text, we can explain the difference in performance between weak and strong measurements in terms of the dynamics of the reservoir. Even with a dynamics that produces states mostly aligned in the z-direction, back-action of weak measurements does not destroy the remaining small quantum coherences of the dynamics. In this way, single-qubit observables in all directions can provide memory. On the other hand, strong measurements in the x and y directions destroy these small coherences, finding that contributions to the memory could become negligible as in the y-direction. 

When dealing with quantum information processing protocols, a central issue is  the possible performance improvement with respect to some meaningful classical limit. In the case of QRC, the most immediate evidence of such an improvement is given by the large size of the Hilbert space with respect to the number of physical nodes. In our model, the classical limit can be identified with the regime where the back-action does not affect the dynamics as all the processing is performed along a fixed direction, with no quantum coherence involved. In this sense, while we have not investigated the classical limit of the model (for instance corresponding to the absence of the external magnetic field),  clear evidence of such an advantage can be found both in recognition and forecasting protocols as all linear independent observables, including noncommuting operators, contribute to give the total processing capability of the system.
\section{Experimental time resources and derivation of equation (\ref{eq:gthr_1})}
\label{appendix_gvaluesandtwo}
If the measurement process takes time $\tau_{\textnormal{M}}$ and the system has to be reset to continue the experiment with an associated preparation time, $\tau_{\textnormal{R}}$, the expressions for the experimental time corresponding to the RSP, the RWP, and the OLP are, respectively:
\begin{align}
\label{texp_restartingAPPENDIX}
t^{\rm RSP}_{\textnormal{exp}}
&=t^{\rm RWP}_{\textnormal{exp}}+3N_{\textnormal{meas}}\left(\frac{1}{2}N_f(N_f+1)\Delta t\right),
\\
\label{texp_rewindingAPPENDIX}
t^{\rm RWP}_{\textnormal{exp}}
&=3N_{\textnormal{meas}}\left(\tau_{\textnormal{wo}}+(\tau_{\textnormal{wo}}+\tau_{\rm R}+\tau_{\rm M})N_f\right),
\\
\label{texp_onlineAPPENDIX}
t^{\rm OLP}_{\textnormal{exp}}
&=3N_{\textnormal{meas}}\left(\tau_{\rm wo}+(\Delta t +\tau_{\rm M}) N_f\right).
\end{align}
The factor $3$ accounts for the need for non-simultaneous measurements for the x, y, and z directions; and $N_f\equiv N_t-N_{\rm wo}$, where we remind that $N_t$ is the total number of time steps $k$ in the input time-series.
In the main text, we have considered that the measurements are done instantaneously and that the time for resetting the system is negligible, namely $\tau_{\rm M}=\tau_{\rm R}=0$.

In the following lines, we provide the details of the derivation of Eq.~\eqref{eq:gthr_1} and the equivalent for two-qubit observables from these last equations. We start with the condition:

\begin{equation}
t_{\textnormal{exp}}^{\textnormal{OLP}} \leq t_{\textnormal{exp}}^{\textnormal{RWP}}.
\end{equation}

By replacing \eqref{texp_restartingAPPENDIX} and \eqref{texp_onlineAPPENDIX} in this last expression, we obtain:

\begin{equation}
\frac{N_{\textnormal{meas}}^{\textnormal{OLP}}}{N_{\textnormal{meas}}^{\textnormal{RWP}}} \leq \frac{\tau_{\textnormal{wo}}+(\tau_{\textnormal{wo}}+\tau_{\textnormal{R}}+\tau_{\textnormal{M}})N_t}{\tau_{\textnormal{wo}}+(\Delta t+\tau_{\textnormal{M}})N_t}.
\end{equation}

In the right part of the inequality, we consider the case when the number of time steps is sufficiently large to neglect the first washout-time term in both the numerator and the denominator. Namely, $N_{\rm wo} \ll N_{t}$. In addition, we assume that the measurement time is the same in both protocols and very fast, i.e. $\tau_{\textnormal{M}} \ll \Delta t$, and that $\tau_{\textnormal{R}} \ll \tau_{\textnormal{wo}}$.
With all these assumptions, we end up with:

\begin{equation}
\frac{N_{\textnormal{meas}}^{\textnormal{OLP}}}{N_{\textnormal{meas}}^{\textnormal{RWP}}} \leq N_{\textnormal{wo}}.
\end{equation}

One main difference between the online and the RWP is the measurement strength used. In the RWP, a projective measurement is the best choice as the back-action effect is irrelevant. However, for the OLP, the measurement strength has to be tuned so that it is sufficiently weak to have a system with exploitable memory but strong enough to extract information and reduce the required number of measurements. Consequently, the number of measurements for achieving the same statistical error in the observables depends on the protocol. The relations between measurements with different measurement strength were derived in Appendix~\ref{Appendix_Measurements_Formalism}, Eqs.~\eqref{eqNmesNmeas} and \eqref{eqNmesNmeastwoqubit}. By using these equations, we find for single-qubit and two-qubit observables, respectively:
\begin{equation}
\frac{N_{\textnormal{meas}}^{\textnormal{OLP}}}{N_{\textnormal{meas}}^{\textnormal{RWP}}}=\frac{(1+g^2)g'^2}{(1+g'^2)g^2},
\end{equation}
\begin{equation}
\frac{N_{\textnormal{meas}}^{\textnormal{OLP}}}{N_{\textnormal{meas}}^{\textnormal{RWP}}}=\frac{(1+2g^2+g^4)g'^4}{(1+2g'^2+g'^4)g^4}.
\end{equation}
$g$ is the measurement strength of the OLP. A projective measurement in the RWP is achieved in the limit of large $g'$, so that the two previous equations are simplified, respectively, to:
\begin{equation}
\frac{N_{\textnormal{meas}}^{\textnormal{OLP}}}{N_{\textnormal{meas}}^{\textnormal{RWP}}}=\frac{1+g^2}{g^2},
\end{equation}
\begin{equation}
\frac{N_{\textnormal{meas}}^{\textnormal{OLP}}}{N_{\textnormal{meas}}^{\textnormal{RWP}}}=\frac{1+2g^2+g^4}{g^4}.
\end{equation}

At this point, we can relate the value of $g$, which is always $g \geq 0$, with the number of washout time steps in order to have the same statistical error with both protocols and to have a faster OLP.

For single-qubit observables, we obtain the condition:
\begin{equation}
\frac{1+g^2}{g^2}\leq N_{\textnormal{wo}},
\end{equation}
which leads to Eq.~\eqref{eq:gthr_1} in the main text:
\begin{equation}
g \geq \sqrt{\frac{1}{N_{\textnormal{wo}}-1}}.
\end{equation}

For two-qubit observables, we have:
\begin{equation}
\frac{1+2g^2+g^4}{g^4}\leq N_{\textnormal{wo}},
\end{equation}
that finally reads as:
\begin{equation}
g \geq \sqrt{\frac{1}{\sqrt{N_{\textnormal{wo}}}-1}}.
\end{equation}
For example, for $N_{\textnormal{wo}}=20$, the measurement strength should be $g  \gtrsim 0.54$.
\providecommand{\noopsort}[1]{}\providecommand{\singleletter}[1]{#1}%
%
\end{document}